\documentclass[12pt,preprint]{aastex}

\begin{document}

\title{ECHO EMISSION FROM DUST SCATTERING AND X-RAY AFTERGLOWS OF GAMMA-RAY BURSTS}

\author{\sc L. Shao\altaffilmark{1,2}, Z. G. Dai\altaffilmark{1}, N. Mirabal\altaffilmark{3}}
\altaffiltext{1}{Department of Astronomy, Nanjing University,
Nanjing 210093, China; lang@nju.edu.cn,
dzg@nju.edu.cn}\altaffiltext{2}{JILA, University of Colorado,
Boulder, CO 80309, USA;
shaol@jilau1.colorado.edu}\altaffiltext{3}{Columbia Astrophysics
Laboratory, Columbia University, New York, NY 10027, USA}

\begin{abstract}
We investigate the effect of X-ray echo emission in gamma-ray bursts
(GRBs). We find that the echo emission can provide an alternative
way of understanding X-ray shallow decays and jet breaks. In
particular, a shallow decay followed by a ``normal" decay and a
further rapid decay of X-ray afterglows can be together explained as
being due to the echo from prompt X-ray emission scattered by dust
grains in a massive wind bubble around a GRB progenitor. We also
introduce an extra temporal break in the X-ray echo emission. By
fitting the afterglow light curves, we can measure the locations of
the massive wind bubbles, which will bring us closer to finding the
mass loss rate, wind velocity, and the age of the progenitors prior
to the GRB explosions.
\end{abstract}

\section{INTRODUCTION}
\label{sec:intro}

Gamma-ray bursts (GRBs) are the most luminous and intriguing
explosions in the universe. A number of physical interpretations for
GRBs have been proposed since they were discovered about four
decades ago. The most successful model for GRBs and their afterglows
is the fireball shock model \citep[and references therein]{m06}.
Although the interpretation for the prompt gamma-ray emission is
still controversial even in this standard scenario, the
long-wavelength afterglows have been widely accepted to be the
emission from relativistic shocks sweeping up an external medium
\citep[and references therein]{m06,z07}.

In spite of the success of the standard external shock model for
interpreting most features of afterglows, there are still a few
issues recently raised by the \textit{Swift} X-ray Telescope (XRT)
\citep{g04}. The first issue is the ubiquitously detected shallow
decay of early X-ray afterglows \citep{n06,z06}. In general, to
account for the shallow decay, a ``refreshed shock" scenario is
assumed, either by a continuous activity of the GRB progenitor
\citep{dl98,zm01,d04}, or by a power-law distribution of the Lorentz
factors in the ejecta \citep{rm98,sm00}. However, although the
``refreshed shock" model is generally able to explain the flatter
temporal X-ray slopes from current analysis, it may not be able to
explain self-consistently both the X-ray and the optical afterglows
of some well-monitored GRBs, such as GRB 050319 and GRB 050401
\citep{fp06}.

The second issue is the lack of jet breaks that are supposed to be
detected simultaneously in X-ray and optical light curves
\citep{p06,s07,br07}. This raises a serious concern about our
understanding of the afterglow emission. Currently, possible
solutions to this problem are additional mechanisms taking place in
the early X-ray emission \citep{fp06}, X-ray emission arising from
different outflows with the optical emission \citep{p06}, or
alternatively, the temporal break in X-rays is masked by some
additional source of X-ray emission \citep{s07}, so that a jet break
at the expected time could still be seen in the optical band. The
last interpretation may bring renaissance to the physical meaning of
a jet break \citep{r99} and $E_{\rm jet}$ \citep{f01}, and therefore
revive the relation between $E_{\rm jet}$ and $E_{\rm peak}$
\citep{g04a}.

Recently \cite{sd07} realized that the impulsive prompt X-ray
emission scattering off the surrounding medium could give rise to a
long-term X-ray flux, which is comparable to that of most observed
afterglows. The scattered X-ray emission (which is called echo
emission herein) could be an additional contributor to the early
X-ray afterglow, which was originally put forward by \cite{k98} and
then further discussed by \cite{mg00} and \cite{ss03}. \cite{sd07}
also worked out detailed light curves of the echo emission that
could nicely reproduce most of the temporal features of observed
X-ray afterglows, i.e., a shallow decay followed by a ``normal"
decay and a further rapid decay \citep{z06,n06}. This scenario
provides a self-consistent solution to both the shallow decay and
the jet break problem. By fitting the X-ray light curves of GRB
060813 and GRB 060814, \cite{sd07} concluded that the scattering
medium, i.e., a dusty shell should be at a distance of about tens of
parsecs from the GRB source.

Coincidentally, this conjectured dusty shell by \cite{sd07} is
consistent with the massive wind bubble around a Wolf-Rayet (WR)
star, which has been proposed as the progenitor of a GRB. The study
of stellar structure and evolution has provided a mass-loss history
of a massive star, starting with the Main Sequence (MS) stage as an
O star, which will evolve into a WR star and then produce a final
supernova and/or a GRB. By the computation of stellar wind and
atmosphere models for a O-type star of 60 ${\rm M_\sun}$ mass,
\cite{l94} concluded that the stellar mass lost to the wind bubble
is about 32 ${\rm M_\sun}$ in the MS stage, 8 ${\rm M_\sun}$ in the
Luminous Blue Variable (LBV) stage, and 16 ${\rm M_\sun}$ in the WR
stage. Such a massive wind bubble will produce a significant shell
mainly swept up by the MS wind at the time of the supernova or GRB
(see Fig.\ref{fig:bubble}), with a distance from the progenitor of
about $R\sim$ tens of parsecs \citep{g96a,m03,d07}. If the clumps
formed during the LBV and WR stages survive long enough, they will
be hitting the MS shell roughly at the time of the supernova or GRB
\citep{g96a}.

In this paper, we apply the echo emission model to a sample of 36
GRBs with well-monitored X-ray light curves by the XRT onboard
\textit{Swift}. We find that most of the light curves without
significant flares can be reproduced by our model. In this sample,
14 bursts have available redshifts by observations. By fitting their
X-ray light curves, we find the distance $R$ of the wind bubbles
from GRB progenitors. A measurement of $R $ might bring us closer to
finding the mass loss rate, density and wind velocity of the
progenitor prior to the explosion.

Our paper is structured as follows: In $\S$\ref{sec:bubble}, we
first discuss the existence of massive stellar wind bubbles around
GRBs when they explodes and the subsequent X-ray echo emission. We
formulate the theoretical light curves of X-ray echo emission in
$\S$\ref{sec:theory}, and we introduce the temporal break in echo
emission and fit the observational light curves of 36 GRBs in
$\S$\ref{sec:break}. Discussion and conclusions follow in
$\S$\ref{sec:discuss}.

\section{THE ECHO ON THE BUBBLE}
\label{sec:bubble}

It has been shown by \cite{sd07} with a detailed calculation
following \cite{m99} that, the fiducial flux of the echo emission
could be comparable to an already observed X-ray afterglow at some
certain time, given that the prompt X-rays from a GRB are
significantly scattered by the circumstellar dust. Here, we provide
a more concise approach by order-of-magnitude estimation.

Evidence accumulated in the last few years has verified the
long-hypothesized origin of soft GRBs in the deaths of WR stars
\citep[and references therein]{wb06}. It is generally believed that
the progenitor WR star will blow out a hot bubble in the
interstellar medium (ISM) by massive stellar wind \citep{g96a,d07}.
The typical radius of the dense shell of swept-up surrounding ISM is
of order tens of parsecs, as given by a self-similar solution
\citep{cmw75,r01a,m03}
\begin{eqnarray}
R&=&\left({25\dot{M}v^2}\over{14\pi n m_{\rm
p}}\right)^{1/5}t^{3/5}\nonumber \\ &=&100\left({\dot{M}_{-6.2}
v_{3.5}^2\over n_0}\right)^{1/5}t_{6.6}^{3/5} \,{\rm pc},
\end{eqnarray}
where the mass-loss rate during the MS stage $\dot{M}=10^{-6.2}\,
{\rm M_\odot/yr}$, the typical wind velocity $v=10^{3.5}$ km/s, and
the lifetime of the MS stage $t=10^{6.6}$ yr are adopted for a 35
${\rm M_\odot}$ star \citep{g96b,d07}, and the number density of
interstellar medium $n=1 \,{\rm cm}^{-3}$ is assumed.

\begin{figure}
\plotone{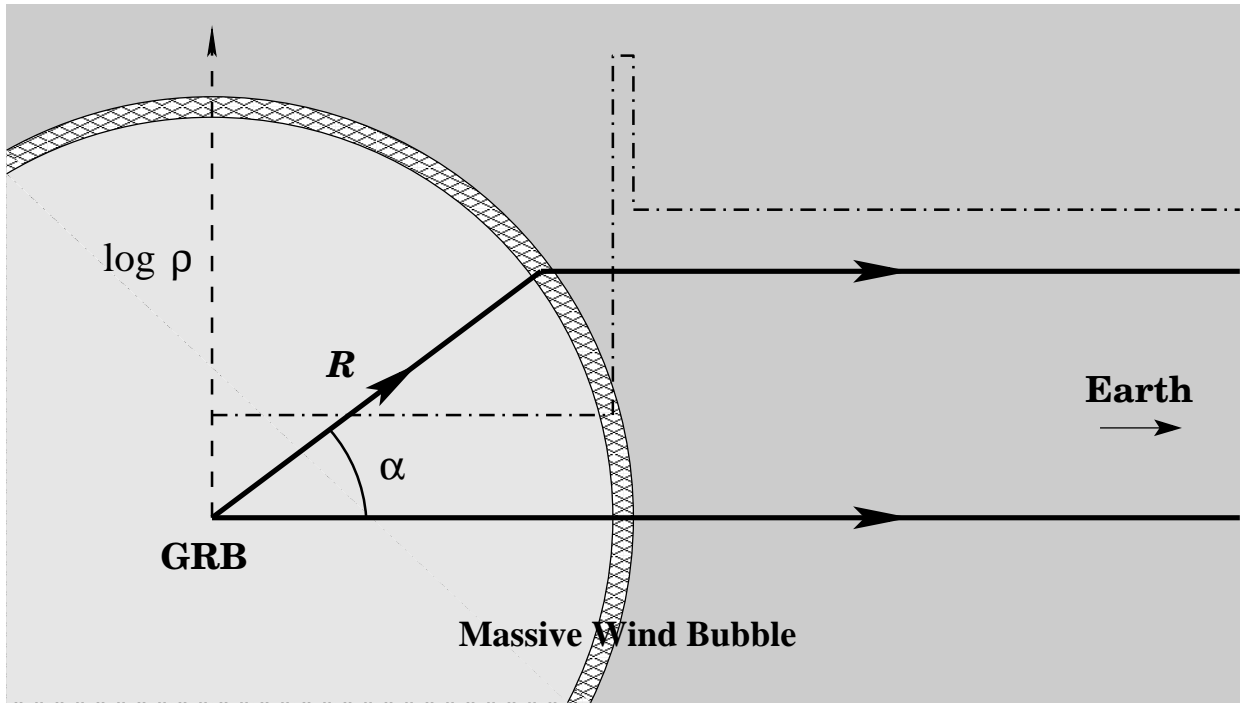} \caption{Geometry and density profile of a massive
wind bubble around a WR star. The major scattering takes place at
the massive and dense shell (marked with brick-wall pattern), which
is a main dusty shell around the WR star (or more probably a WC
star). Inside the shell, the interior structure of the bubble
(filled with light gray) is ignored, which is mainly determined by
the LBV and WR winds and not important for scattering, since the
density is much lower there and essentially no dust survives close
to the WR star. Outside the shell, the interstellar medium (ISM)
dominates (filled with heavy gray). Meanwhile, the gas density
$\rho$ is also plotted with logarithmic scale in dot-dashed line
\citep[not in scale, see][]{g96a,d07}.} \label{fig:bubble}
\end{figure}

Correspondingly, the geometry and density profile of a massive wind
bubble around a WR star are shown in Fig.\ref{fig:bubble}. WR stars
-- characterized by high mass-loss rate -- are believed to form
massive dust in their winds \citep[e.g., ][]{w87,c03,c07a}.
Interestingly, all dust shells that were observed were essentially
associated with carbon-rich WR stars of the latest subtypes of the
WC sequence only \citep[which are called WC stars herein, e.g.,
][]{a72,gh74}. WC stars are the final evolutionary phase of the most
massive ($\geq 40 {\rm M_\sun}$) stars, which, in some sense seem to
be favored here as the progenitors of some GRBs, since there is
where echo emission would be most efficient.

Assuming that an X-ray burst (i.e., the X-ray counterpart of a
prompt GRB produced at the final stage of a WC star) with a fluence
$S_{\rm X}$ is scattered by the dusty shell at a distance $R$ with a
scattering optical depth $\tau_{\rm sca}$, the echo emission is
spread within a characteristic time scale $t_{\rm d}$, due to the
scattering delay of X-rays with different scattering angles $\alpha$
(see Fig.\ref{fig:bubble} for the geometry of scattering). Since the
characteristic scattering angle $\alpha$ is physically small
\citep{ah78}, the time scale $t_{\rm d}$ is consequently much
smaller than $R/c$, where $c$ is the speed of light. Therefore, we
have a fiducial flux of the X-ray echo emission over a time $T_{\rm
d}$
\begin{eqnarray}
F^{\rm echo}&=&{S_{\rm X}  \times \tau_{\rm sca}\over t_{\rm d}} \nonumber \\
&=&2.5\times10^{-11}\,\left({S_{\rm X}\over2.5\times10^{-7}\,{\rm
ergs\,cm^{-2}}}\right)\left({\tau_{\rm sca}\over
0.1}\right)\left({t_{\rm d}\over 10^3\,{\rm s}}\right)^{-1} \,{\rm
ergs\,cm^{-2}\,s^{-1}}, \label{eq:flux}
\end{eqnarray}
where $S_{\rm X}$ is the total fluence of the burst observed in
X-ray band (say, 0.3-10 keV), the echo flux and the source flux are
assumed to suffer the same interstellar absorption on their way to
the observer, and the cosmological $k$-correction is ignored.
$S_{[0.3-10]}=2.5\times 10^{-7}\,{\rm ergs\,cm^{-2}}$ is
approximated here, assuming $S_{[0.3-10]}/S_{[15-150]}= 1/4$ and
$S_{[15-150]}=10^{-6}\,{\rm ergs\,cm^{-2}}$, given that
$S_{[50-100]}/S_{[25-50]}\sim 1$ \citep{s07a} and most GRBs have a
flat spectrum in soft X-ray band \citep{p00,s07a}. Note that, a
realistic $S_{[0.3-10]}$ should be essentially larger than the value
we adopt here, because the X-rays between [0.3-10] keV before and
after the BAT trigger interval \citep[$T_{100; }$][]{s07a},
especially those from the soft tail emission in most GRBs, are not
included in this approximation. Since the scattering angle $\alpha$
is very small, we have
\begin{eqnarray}
t_{\rm d}&=&{(1+z)R \alpha^2\over 2c} \nonumber \\
&=&10^3\,\left(1+z\over 2\right)\left(R\over 100\,{\rm
pc}\right)\left({\alpha\over
 1\arcmin}\right)^2\,{\rm s},
 \label{eq:timescale}
\end{eqnarray}
where $z$ is the redshift of the GRB, and $\alpha$ is the typical
scattering angle of X-ray echo emission \citep{ah78,k98}.

Hence, we have three interesting coincidences:
\begin{enumerate}
\item
The dusty shell at a distance $R$ of about tens of parsecs assumed
by \cite{sd07} is consistent with the massive wind bubble around a
WR star at its final stage \citep{g96a,m03,d07};
\item
The timescale of the X-ray echo emission on such a wind bubble is
about $t_{\rm d} \sim 10^3$ s, which is the typical timescale when
the shallow decay phase dominates in X-ray afterglow
\citep{z06,n06};
\item
The X-ray echo flux $F^{\rm echo}$ from such a wind bubble is
consistent with most of both pre-\textit{Swift} and \textit{Swift}
X-ray afterglows around $10^3$ s \citep[e.g.,][]{c99,n06}.
\end{enumerate}

These three coincidences motivate the study of echo emission as an
additional contributor to the X-ray emission. In what follows, we
work out the light curve of the echo emission, and show the fourth
and the most spectacular coincidence that the theoretical light
curves of X-ray afterglows are consistent with the observed ones.

\section{THEORETICAL LIGHT CURVE}
\label{sec:theory}

\cite{sd07} have derived the temporal profile of the echo emission.
They showed that the feature in the X-ray light curve of a shallow
decay followed by a ``normal" decay and a further steepening could
be reproduced, given that the grains in the dusty shell have a
distribution over size and that the scattering is dependent on the
wavelength of X-rays through the Rayleigh-Gans approximation.
According to \cite{sd07}, this theoretical light curve can be
formulated as
\begin{eqnarray}
F^{\rm echo}(t)=A \int_{E_-}^{E_+}\int_{a_-}^{a_+}
S(E)\tau(E,a)j_1^2\{x[\alpha(t)]\}{1\over t} {\rm d}a{\rm d}E,
\label{eq:lightcurve}
\end{eqnarray}
where $A$ is a constant, $E$ is the energy of an X-ray photon, $a$
is the size of a grain in the dusty shell, and we define the
following functions
\begin{equation}
S(E)=\left({E\over{100\,{\rm keV}}}\right)^\delta {\rm
exp}\left[-{(\delta+1)E\over E_{\rm p}}\right], \label{eq:s}
\end{equation}

\begin{equation}
\tau(E,a)=\left[{(1+z)E\over{1\,{\rm
keV}}}\right]^{-s}\left({a\over0.1\,\micron}\right)^{4-q},
\end{equation}

\begin{equation}
j_1(x)={\sin x\over x^2}- {\cos x\over x},
\end{equation}

\begin{equation}
x(\alpha)= {2\pi (1+z)E a \alpha\over hc},
\end{equation}

\begin{equation}
\alpha(t)=\sqrt{2ct\over(1+z)R}, \label{eq:alpha}
\end{equation}
where $\delta$ and $E_{\rm p}$ are the spectral index and the
observed peak energy of the source \citep{p00}, $s$ and $q$ are the
power-law indices in the formula of scattering optical depth
\citep{mg86,m77,sd07}, $a_-$ and $a_+$ are the cutoff sizes of dust
grains \citep{m77}, $[E_-,E_+]$ is the energy band of the detector,
and $h$ is the Planck constant (see the Appendix for a detailed
derivation).

\section{TEMPORAL BREAK IN X-RAY ECHO EMISSION}
\label{sec:break}

Thanks to the \textit{Swift} satellite \citep{g04}, our
interpretation of GRBs has become more complex. The feature of a
shallow decay followed by a ``normal" decay and a further rapid
decay of X-ray afterglows appears to be canonical \citep{z06,n06}.
Coincidentally, we find that this feature can be naturally
reproduced by the echo emission.

Well-monitored X-ray light curves of 36 GRBs (see Fig.
\ref{fig:lightcurve}) were fitted by our theoretical model above
(see Eq. \ref{eq:lightcurve}), using the online data by \cite{e07}.
A total of 14 events have known redshifts\footnote{ The redshift
data come from \url{http://www.mpe.mpg.de/$\sim$jcg/grbgen.html}}
(see Tab. \ref{tab:redshift}) while 22 without known redshifts are
fixed to be at $z=1$ (see Tab. \ref{tab:noredshift}). We assume
$E_-=0.3$ keV and $E_+=10$ keV for \textit{Swift} XRT
\citep{g04,e07}, $a_-=0.005\, \micron$ \citep{m77}, $\delta=0$, and
$E_{\rm p}=200$ keV \citep{p00}. The other parameters, $s$, $q$,
$a_+$, and $R$ are given in Tab. \ref{tab:redshift} and Tab.
\ref{tab:noredshift}. Apparently, $a_+$ is almost a constant here
(i.e., $a_+=0.25\, \micron$, except $a_+=0.5\, \micron$ for GRB
050505) and consistent with most results from optical observations
\citep{m77,ps95}.

Here $A$ is a numerical coefficient, which is equal to 2$A_1 A_2$ as
shown in the Appendix. $A_1$ is the coefficient in the integrated
GRB spectrum between 0.3 and 10 keV, which is not available based on
current working frequency of BAT onboard \textit{Swift}. Similarly,
$A_2$ is hardly determined by current observations about the X-ray
optical depth. A further analysis of the parameter $A_1$ and $A_2$
is required once the relevant observational data are sufficient.

In Fig. \ref{fig:lightcurve}, all the light curves approach to a
steep decay ($\propto t^{-2}$), which is already predicted by Eq.
(\ref{eq:lightcurve}). Since this is a double integral about $E$ and
$a$, $t$ is an independent parameter. Approximately, when
$x[\alpha(t)]\gg1$, we have
\begin{eqnarray}
F^{\rm echo}(t) &\propto& F_{E,a}^{\rm echo}(t) \nonumber \\
&\propto& \left\{ {\sin x[\alpha(t)]\over x^2[\alpha(t)]}- {\cos
x[\alpha(t)]\over x[\alpha(t)]}\right\}^2{1\over t} \nonumber \\
&\propto& {\cos^2 x[\alpha(t)]\over x^2[\alpha(t)] t} \nonumber \\
&\propto&  t^{-2},
\end{eqnarray}
where $\cos^2 x[\alpha(t)]$ is a high-frequency periodic function of
$t$ when $x[\alpha(t)]\gg1$, and will be independent on $t$ after
the integration over $E$ and $a$. Therefore, letting
$x[\alpha(t)]=1$, we have a critical time
\begin{eqnarray}
t_{\rm c}&=&{h^2 c R \over 8 \pi^2 (1+z) a^2 E^2} \nonumber \\
&=& 10^4s \left({2\over1+z}\right)\left(a \over 0.1 \micron
\right)^{-2} \left({E} \over 1{\rm keV}\right)^{-2}\left({R \over
100 {\rm pc}}\right), \label{eq:td}
\end{eqnarray}
after which, the light curves will be approaching the steep decay
($\propto t^{-2}$).

By fitting the light curves, we can indirectly derive the distant
$R$ for a given event at redshift $z$ (see Table
\ref{tab:redshift}). These values of $R$ appear to be consistent
with the observed radii of wind bubbles around WC stars
\citep{m97,c99a}. For bursts without known redshifts we only list a
pesudo value of the bubble radius $R_{\rm pseudo}$ assuming a fixed
redshift $z=1$ (see Table \ref{tab:noredshift}). The larger scatter
in $R_{\rm pseudo}$ values for the latter group appears to be a
direct consequence of the arbitrary $z=1$ prescription.

\section{DISCUSSION AND CONCLUSIONS}
\label{sec:discuss}

Recently, \cite{l07a} provided a comprehensive analysis of shallow
decays and ``normal'' decay segments in GRBs. They concluded in
their sample that most GRBs in these two segments have a spectral
index $\beta$ around $-1.1$ (as the flux density $F_{E}\propto
E^{\beta}$). Here we can also get a knowledge of the spectrum of
echo emission based on our analysis (see Eq. \ref{eq:fluxdensity}).
Approximately, when $x[\alpha(t)]\simeq \pi$ at a given time
$t_{\pi}$, we have

\begin{eqnarray}
F_{E}^{\rm echo}(t_{\pi}) &\propto& F_{E,a}^{\rm echo}(t_{\pi})
\nonumber \\ &\propto& S(E)\tau(E,a) \nonumber \\
&\propto& E^{\delta-s} \,,
\end{eqnarray}
where $t_{\pi}$ is about an order of magnitude larger than $t_{\rm
d}$ in Eq. (\ref{eq:td}), and roughly equals most of the $t_2$'s
defined in \cite{l07a}. Therefore, we should have the spectral index
$\beta \gtrsim \delta-s$ before $t_{\pi}$, since the spectrum of
echo emission is softening in the long run. As shown in Fig.
\ref{fig:hr}, the hardness ratio of echo emission is monotonously
decreasing. Interestingly, it reaches a platform \citep[instead of
an extremum proposed by ][]{k98} almost around $t_{\pi}$ and then
keeps decreasing very slowly. If we assume $\delta \simeq 0$
\citep{p00} and $s \sim 2$ \citep{mg86,mg00}, we should have $\beta
\gtrsim -2$. This seems to be inconsistent with the observational
result by \cite{l07a}, since they ended up with $\beta \sim -1.1$.

However, we have two reasons that lead us to be content with our
result based on current analysis. First,  since we are dealing with
the prompt X-ray emission and the $\delta \simeq 0$ here is the
result from hard X-rays by BATSE \citep{p00}, we can imagine that
the real average $\delta$ should be larger than 0. Specifically,
based on the standard synchrotron emission, $\delta$ should equal
$1/3$, and the synchrotron self-absorption leads to $\delta \sim 2$.
Second, the Rayleigh-Gans approximation is adopted here to calculate
the differential cross section of scattering. This approximation
overestimates the echo emission at softer X-ray band (say, in
0.3-1.5 keV), where absorption is important, which will change the
spectral shape in soft X-ray band and alleviate the dilemma of a
dramatically small hardness ratio in Fig. \ref{fig:hr}. Based on the
current analysis, it is safe to say that -2 is only a lower limit
for $\beta$. Therefore, our result here is generally consistent with
the observations. Thus, the echo emission cannot be completely ruled
out for the reason as discussed by \cite{l07a}.

Originally, the optical echo emission was used to explain the red
bumps observed in some optical afterglows \citep{eb00}. However,
since the typical optical/infrared scattering angle would be about
$\alpha \sim 60 \degr$ \citep{w79}, the timescale in
Eq.(\ref{eq:timescale}) would be longer, and the optical/infrared
echo emission would be much weaker, and hard to be detected
\citep{r01,h07}. Nevertheless, if the WR wind within a much shorter
distance ($R \sim 10^{14}-10^{15}$ cm) is also taken into account,
there might be detectable variations in the optical afterglows due
to echo emission \citep{mr05}. However, the average LBV or WR wind
bubble around a massive star at its death would be much larger
\citep{gm95a,gm95b}. Therefore, the features of echo emission in the
optical band would be not significant enough, which is possibly the
reason that the standard external shock model was more successful in
the pre-\textit{Swift} era, when the optical observations dominated.

Furthermore, the dramatic variations in the LBV or/and WR winds
would drive instabilities that produce radial filaments in the LBV
or WR shells \citep{g96a,g96b,d07}, which could be the origin of
some flares in the X-rays afterglows \citep{b06}. Therefore, there
might be another timescale due to the angular variation of these
filaments, which could be given by
\begin{eqnarray}
t_{\rm f}&=&{(1+z)R_{\rm f} \alpha_{\rm f}^2\over 2c},
\end{eqnarray}
where $R_{\rm f}$ is the distance of the filaments from the
progenitor and $\alpha_{\rm f}$ is the typical angular size of the
clumps and ripples in the filaments.  To account for the mean ratio
of the width and peak time of the X-ray flares \citep{b07,c07},
i.e., $t_{\rm f}/t_{\rm d} \sim <\Delta t / t> \sim 0.1$,  we may
have $\alpha_{\rm f} \sim 0.1 \,\alpha \sim 0.1 \arcmin$.

\cite{ss03} claimed on a different scattering circumstance
(molecular and atomic matter instead of dust grains) that the
opening angle of jets might be an important factor to the echo light
curve, based on the popular assumption that the GRB prompt emission
is beamed. Since only small-angle scattering ($\alpha \sim
1\arcmin$) for dust grains is considered here, the typical jet
properly pointing at us with a much larger opening angle of about
few degrees will not affect out results. However, the GRB jet might
be pointing at us with its edge in some case, where our result here
would not be valid. Instead, a rapid drop that is much steeper than
$t^{-2}$ will emerge at a late time \citep{ss03}. This is probably
the case for some GRBs, e.g., GRB 060526 \citep{dx07}.

In this paper, we have presented the X-ray echo emission in GRB
phenomenon and we can draw the following conclusions:
\begin{enumerate}
\item
The shallow decay phase can still be understood without additional
contributions from the central engine \citep{z06,l07a}.
Interestingly, the feature of a shallow decay followed by a
``normal" decay and a further rapid decay can be reproduced by the
echo from the prompt X-ray emission scattered by the massive wind
bubble around the progenitor \citep{g96a,m03,d07}. In general, dust
formation seems to be prevailing in carbon-rich WR stars of the WC
subtype and there is where echo emission would be most efficient.

\item
Echo emission can be an additional X-ray source that masked the
X-ray jet breaks \citep{s07}. Therefore, the lack of jet breaks
\citep{p06,br07,p07,l07b} might also be understood via competing
echo emission. In addition, some results in the pre-\textit{Swift}
era, say, some correlations relevant for optical breaks might still
be valid \citep{f01,g04a,lz05} .

\item
There is an extra temporal break in X-ray echo emission introduced
by $t_{\rm c}$ above. By fitting the light curves, we can indirectly
measure the locations of the massive wind bubbles,  which will bring
us closer to finding the mass loss rate, wind velocity, and the age
of the progenitors prior to the GRB explosions \citep{g96a,d07}.

\item
Even though some features in X-ray afterglows would be better
understood if the echo emission is taken into account, the external
shock model is still the dominant scenario for optical/infrared
afterglows, since echo emission is negligible in the later. In
addition, the echo emission model does not completely rule out other
interpretations (energy injection, etc.) in the framework of the
fireball-shock model. Technically, echo emission takes place at a
distance of $\sim$ tens of parsecs, while the internal or/and
external shocks are believed to take place within a distance of
$\sim 10^{17}$ cm. It is more likely that both of them are taking
place in the X-ray afterglow, which potentially complicates the
light curves. However, this may raise the problem about GRB
radiative efficiencies again. Even by assuming that the X-ray
emission is the forward shock emission, the GRB efficiency is
already too high to be accommodated by the leading internal shock
models \citep[and references therein]{z07a}. Assuming an
echo-dominated X-ray emission would exacerbate the situation.
\end{enumerate}

\acknowledgments We are grateful to the referee for constructive
comments and suggestions. This work made use of data supplied by the
UK \textit{Swift} Science Data Centre at the University of
Leicester. We would like to thank Bing Zhang for helpful discussions
and comments. L.S. would also like to thank Rosalba Perna for kindly
hosting during his study in University of Colorado at Boulder. This
work is supported by the National Natural Science Foundation of
China (grants 10221001 and 10640420144) and the National Basic
Research Program of China (973 program) No. 2007CB815404. L.S. was
also supported by the Scientific Research Foundation of Graduate
School of Nanjing University and the State Scholarship Fund by China
Scholarship Council.

\appendix
\section{APPENDIX: DERIVATION OF THE THEORETICAL LIGHT CURVE}
The single-scattering approximation is adopted, which is valid up to
a scattering optical depth, $\tau_{\rm sca}$, of $\sim$0.5 at 1 keV
\citep{pk96,k98}. Therefore, the echo intensity at the Earth per
unit photon energy per unit size of the grains is \citep[see Fig. 1
of ][ for the geometry and quantities defined in this
Appendix]{sd07}
\begin{eqnarray}
I_{E,a}^{\rm echo}(\theta,\phi;t)&=&F_{E}[t-t_{\rm d}(\theta)]\tau_a
(E){{\rm d}\sigma\over \sigma_{\rm sca} {\rm d}\Omega},
\label{eq:intensity}
\end{eqnarray}
where $F_{E}[t-t_{\rm d}(\theta)]$ is the flux density of the
initial GRB before a time delay $t_{\rm d}(\theta)$, which is also
dependent on the viewing angle $\theta$,  $\tau_a(E)$ is the
frequency-dependent scattering optical depth per unit size of the
grains, and the last term is the fraction of the scattered photons,
which go into the line of sight in a unit solid angle. The zero time
point is set as the trigger of a GRB on the Earth. $\phi$ is the
azimuthal angle along the line of sight.

Since the initial GRB is usually much shorter than the long-term
afterglow, we can approximate the initial GRB light curve as a
pulse. We also assume that the initial GRB spectrum is given by a
Band spectrum in the X-ray band \citep{b93}, so we have
\begin{eqnarray}
F_{E}[t-t_{\rm d}(\theta)]&=&A_1\left({E\over{100\,{\rm keV}}}
\right)^\delta {\rm exp}\left[-{(\delta+1)E\over E_{\rm
p}}\right]\times \delta[t-t_{\rm d}(\theta)],
\end{eqnarray}
where we only need the segment of the Band spectrum in the lower
energy band. $A_1$ is a constant, and the time delay from a certain
viewing angle $\theta$ is given by \citep{m99}
\begin{eqnarray}
t_{\rm d}(\theta)={(1+z)D_{\rm d} D_{\rm s} \theta^2 \over{2 c
D_{\rm ds}}}\,.
\end{eqnarray}
Here the Dirac delta function of $t$ can be translated into a
function of $\theta$, using the special properties of its own,
\begin{eqnarray}
\delta[t-t_{\rm d}(\theta)] = {c D_{\rm ds}\over (1+z)D_{\rm d}
D_{\rm s} \theta}\delta[\theta-\hat{\theta}(t)],
\end{eqnarray}
where the function $\hat{\theta}(t)$ is defined as
$\hat{\theta}(t)\equiv[2ctD_{\rm ds}/(1+z)D_{\rm d}D_{\rm s}]^{1/2}$
\citep[see the Appendix of ][]{sd07}.

We also assume the frequency-dependent scattering optical depth per
unit size of the grains is
\begin{eqnarray}
 \tau_a(E)&=&A_2\left[{(1+z)E\over{1\,{\rm keV}}}\right]^
 {-s}\left({a\over0.1\,\micron}\right)^{4-q} ,
\end{eqnarray}
where $A_2$ is a constant with a dimension of $[a^{-1}]$. This is
valid only in a range of the size of the grains, say,
$a\in[a_-,a_+]$ \citep[and the references therein]{sd07}.

For the last term in Eq. (\ref{eq:intensity}), we have $d
\Omega=\theta {\rm d}\theta {\rm d}\phi$, since $\theta$ is very
small. We can do some rearrangement
\begin{eqnarray}
{{\rm d}\sigma\over \sigma_{\rm sca} {\rm d}\Omega}&=&{{\rm
d}\sigma\over \sigma_{\rm sca} {\rm d}\Omega_{\rm SC}}{{\rm
d}\Omega_{\rm SC}\over {\rm d }\Omega},
\end{eqnarray}
where $\sigma_{\rm sca}$ is the total scattering cross-section,
${\rm d}\Omega_{\rm SC}=\alpha {\rm d}\alpha {\rm d} \phi$ is
defined since $\alpha$ is also very small, and we have
$\alpha=(D_{\rm s}/D_{\rm ds})\theta$ in geometry. Neglecting the
chemical composition or shape of the grains, the Rayleigh-Gans
approximation is valid for the differential cross-section
\citep{o65,ah78}
\begin{eqnarray}
{{\rm d}\sigma\over \sigma_{\rm sca}{\rm d}\Omega_{\rm
SC}}&=&{2\over \pi\ }{j_1^2(x)\over \alpha^2},
\end{eqnarray}
where $x\equiv2\pi(1+z)aE\alpha/hc$, and $j_1(x)=({\rm sin}\,
x)/x^2-({\rm cos}\, x)/x$ is the first-order spherical Bessel
function.

Since the wind bubble is very close to the GRB, we have $D_{\rm
s}\simeq D_{\rm d}$ and $D_{\rm ds}\equiv R$. Therefore, the echo
flux density at the Earth per unit size of the grains is
\begin{eqnarray}
F_{E,a}^{\rm echo}(t)&=&\int I_{E,a}^{\rm echo}(\theta,\phi;t) {\rm cos}\,\theta{\rm d}\Omega\nonumber\\
&=&A S(E)\tau(E,a)j_1^2\{x[\alpha(t)]\}{1\over t},
\end{eqnarray}
where A is a numerical constant, and $S(E)$, $\tau(E,a)$, $j_1(x)$,
$x(\alpha)$, and $\alpha(t)$ are given by Eqs.
(\ref{eq:s})-(\ref{eq:alpha}). So, the flux density and the flux of
echo emission at Earth will be
\begin{eqnarray}
F_{E}^{\rm echo}(t)&=&\int_{a_-}^{a_+} F_{E,a}^{\rm echo}(t) {\rm
d}a \label{eq:fluxdensity}
\end{eqnarray}
and
\begin{eqnarray}
F^{\rm echo}(t)&=&\int_{E_-}^{E_+}F_{E}^{\rm echo}(t){\rm d}E \,,
\end{eqnarray}
respectively.

\begin{deluxetable}{lcccccccc}
\tablecolumns{9} \tablewidth{0pc} \tablecaption{Model parameters of
14 GRBs with known redshifts.} \tablehead{ \colhead{GRB} &
\colhead{z\tablenotemark{a}} & \colhead{Begin (s)\tablenotemark{b}}
& \colhead{End (s)\tablenotemark{c}} & \colhead{s} & \colhead{q} &
\colhead{$a_+ (\micron)$} & \colhead{$R $(pc)} &
\colhead{$\chi^2/{\rm dof}$}} \startdata
GRB 050319  & 3.240 & $4\times10^2$ & $2\times10^6$& 2.5 & 4.4 & 0.25 &75 & 1.72 \\
GRB 050505  & 4.27  & $3\times10^3$ & $2\times10^6$& 2.0 & 4.0 & 0.5 &50 & 1.42 \\
GRB 050814  & 5.3   & $1\times10^3$ & $1\times10^6$& 3.0 & 4.5 & 0.25 &55 & 3.53 \\
GRB 051022  & 0.8   & $1\times10^4$ & $1\times10^6$& 2.0 & 3.1 & 0.25 &30 & 2.85 \\
GRB 060210  & 3.91  & $3\times10^3$ & $8\times10^5$& 2.0 & 3.0 &0.25 &200& 1.78 \\
GRB 060218  & 0.033 & $1\times10^4$ & $1\times10^6$& 3.0 & 3.1 & 0.25&50 & 1.92 \\
GRB 060502A & 1.51  & $4\times10^3$ & $2\times10^6$& 3.0 & 4.3 & 0.25&65 & 1.26 \\
GRB 060512  & 0.4428& $3\times10^3$ & $3\times10^5$& 2.0 & 3.0 & 0.25&45 & 2.63 \\
GRB 060707  & 3.425 & $1\times10^3$ & $2\times10^6$& 3.0 & 4.5 & 0.25&170& 2.53 \\
GRB 060714  & 2.711 & $3\times10^2$ & $1\times10^6$& 2.0 & 4.0 & 0.25&95 & 2.61 \\
GRB 060814  & 0.84 & $8\times10^2$ & $1\times10^6$& 2.0 & 4.0 &0.25 & 34 & 2.33 \\
GRB 060729  & 0.54  & $5\times10^2$ & $1\times10^7$& 3.0 & 5.5 & 0.25&45 & 2.93 \\
GRB 061110A & 0.758 & $6\times10^3$ & $7\times10^5$& 3.0 & 4.5 &0.25 & 76 & 2.61 \\
GRB 061121  & 1.314 & $2\times10^2$ & $2\times10^6$& 2.0 & 4.0 & 0.25&40 & 2.30 \\
\enddata
\label{tab:redshift} \
\tablenotetext{a}{\url{http://www.mpe.mpg.de/$\sim$jcg/grbgen.html}}
\tablenotetext{b}{Beginning time of the data for fitting}
\tablenotetext{c}{Ending time of the data for fitting}
\end{deluxetable}

\begin{deluxetable}{lcccccccc}
\tablecolumns{8} \tablewidth{0pc} \tablecaption{Model parameters of
22 GRBs without known redshifts\tablenotemark{a}} \tablehead{
\colhead{GRB} &  \colhead{Begin (s)\tablenotemark{b}} & \colhead{End
(s)\tablenotemark{c}} & \colhead{s} & \colhead{q} & \colhead{$a_+
(\micron)$} & \colhead{$R_{\rm pseudo}$ (pc)}    &
\colhead{$\chi^2/{\rm dof}$}} \startdata
GRB 050712  & $4\times10^3$ & $2\times10^6$& 2.3 & 4.3 &0.25 & 55 & 2.75 \\
GRB 050713A & $4\times10^3$ & $2\times10^6$& 2.0 & 4.0&0.25  & 45 & 2.67 \\
GRB 050713B & $5\times10^2$ & $1\times10^6$& 3.0 & 4.5 &0.25 & 70 & 2.21 \\
GRB 050802  & $3\times10^2$ & $1\times10^6$& 2.0 & 3.1 &0.25 & 30 & 4.64 \\
GRB 050822  & $1\times10^3$ & $5\times10^6$& 3.0 & 4.5 &0.25 & 80 & 2.19 \\
GRB 050915B & $1\times10^3$ & $5\times10^5$& 3.0 & 4.8 &0.25 & 70 & 1.60 \\
GRB 051008  & $3\times10^3$ & $4\times10^5$& 2.5 & 3.5 &0.25 & 10 & 1.93 \\
GRB 051016A & $4\times10^3$ & $6\times10^5$& 2.3 & 4.0 &0.25& 80 & 6.44 \\
GRB 060204B & $4\times10^2$ & $4\times10^5$& 2.0 & 3.1 &0.25 & 25 & 1.64 \\
GRB 060219  & $2\times10^2$ & $4\times10^5$& 2.5 & 3.5 &0.25 & 45 & 1.04 \\
GRB 060306  & $3\times10^2$ & $3\times10^5$& 2.0 & 4.0 &0.25 & 50 & 1.41 \\
GRB 060428A & $3\times10^2$ & $3\times10^6$& 2.7 & 4.2 &0.25 & 95 & 2.26 \\
GRB 060510A & $1\times10^2$ & $5\times10^5$& 3.0 & 4.5 &0.25 & 10 & 2.28 \\
GRB 060604  & $3\times10^3$ & $1\times10^6$& 2.0 & 4.0 &0.25 & 80 & 1.86 \\
GRB 060708  & $2\times10^2$ & $1\times10^6$& 2.0 & 4.0 &0.25 & 50 & 2.03 \\
GRB 060813  & $9\times10^1$ & $2\times10^5$& 2.0 & 3.1 &0.25 & 10 & 2.68 \\
GRB 061004  & $3\times10^2$ & $1\times10^5$& 2.0 & 4.0 &0.25 & 15 & 2.61 \\
GRB 061222A & $2\times10^2$ & $2\times10^6$& 2.1 & 4.2 &0.25 & 30 & 1.90 \\
GRB 070129  & $1\times10^3$ & $2\times10^6$& 3.0 & 4.5 &0.25 & 90 & 1.63 \\
GRB 070328  & $1\times10^2$ & $6\times10^5$& 1.9 & 4.0 &0.25 & 5 & 4.75 \\
GRB 070419B & $4\times10^3$ & $4\times10^5$& 2.0 & 3.1&0.25  & 20 & 2.74 \\
GRB 070420  & $3\times10^2$ & $4\times10^5$& 3.0 & 3.1 &0.25 & 15 & 3.07 \\
\enddata
\label{tab:noredshift}
\tablenotetext{a}{We assume z=1 to carry out the fitting}
\tablenotetext{b}{Beginning time of the data for fitting}
\tablenotetext{c}{Ending time of the data for fitting}
\end{deluxetable}

\begin{figure}
\plottwo{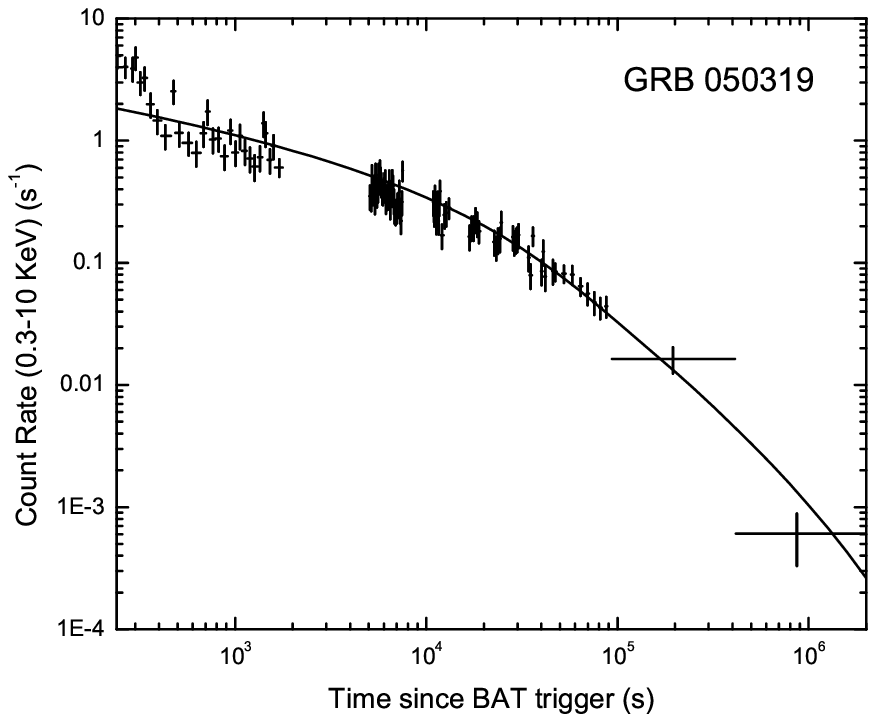}{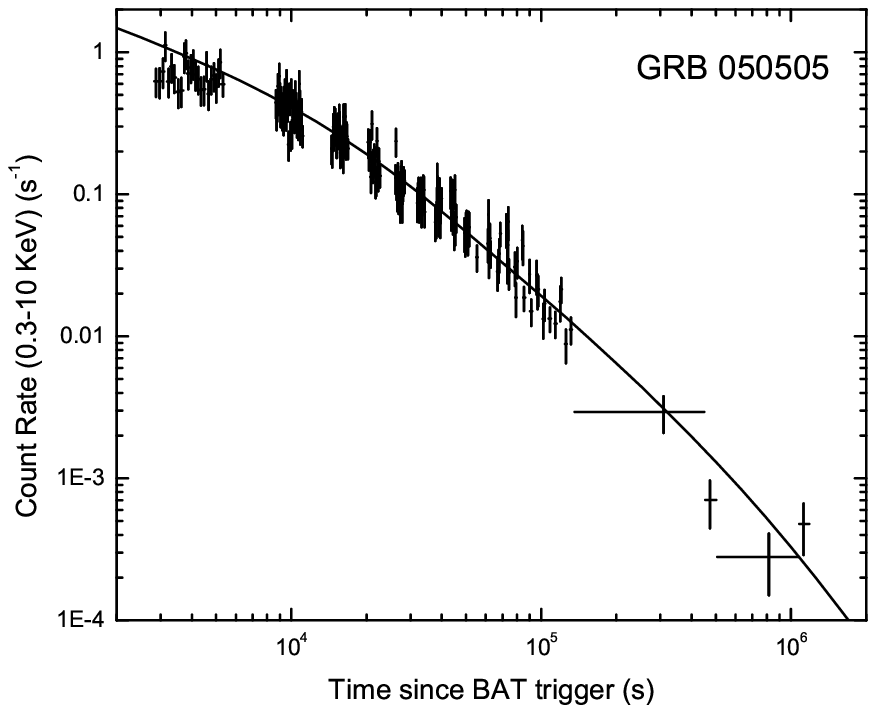}
\plottwo{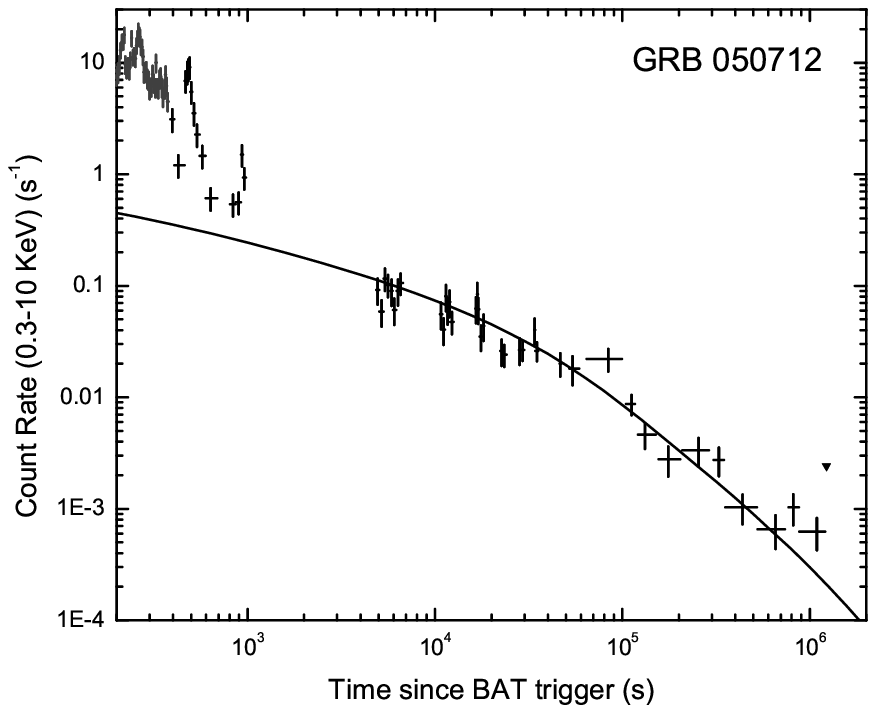}{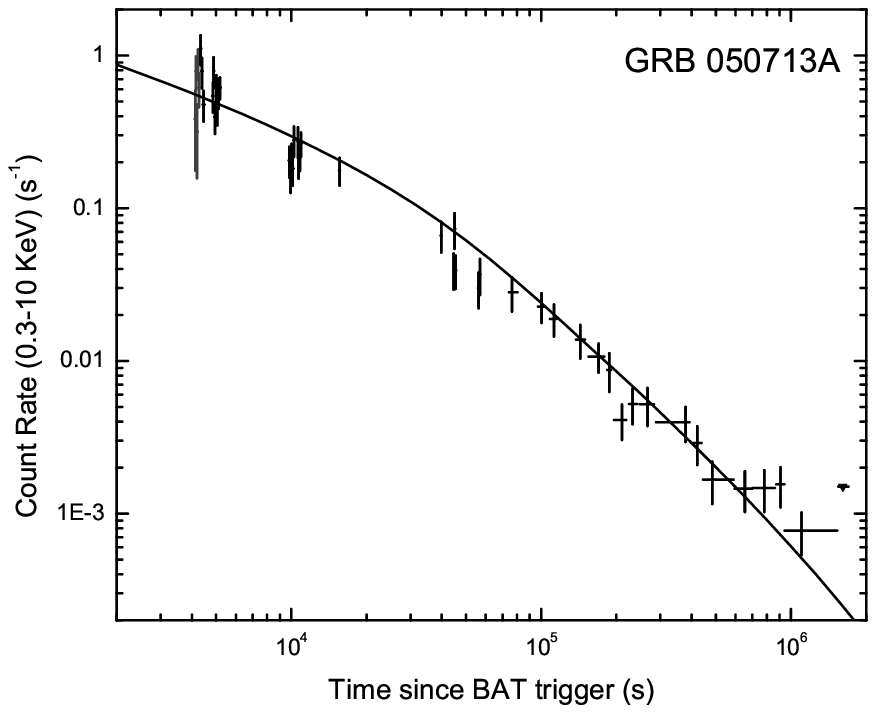}
\plottwo{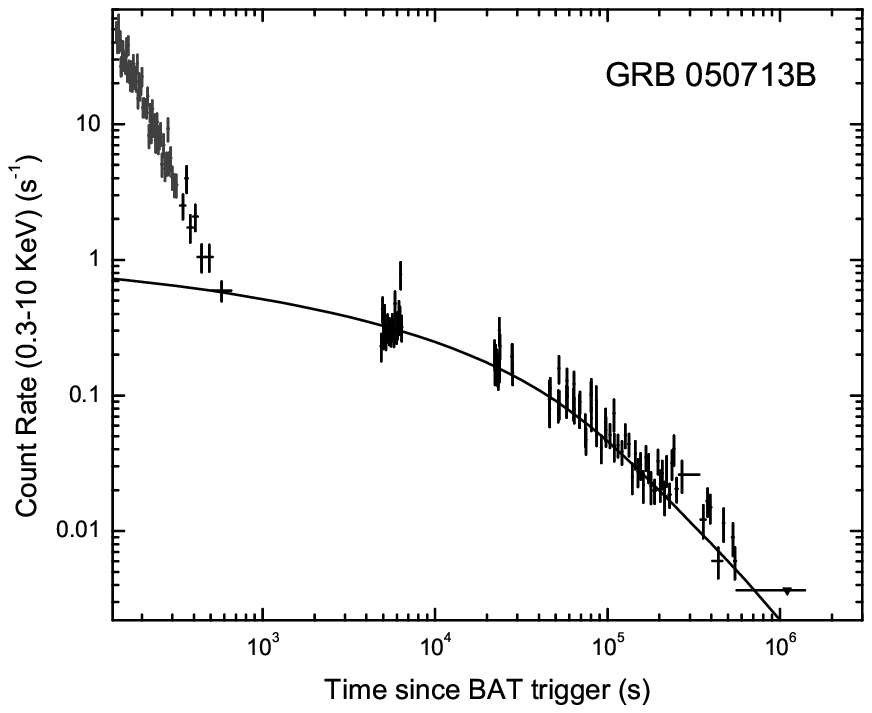}{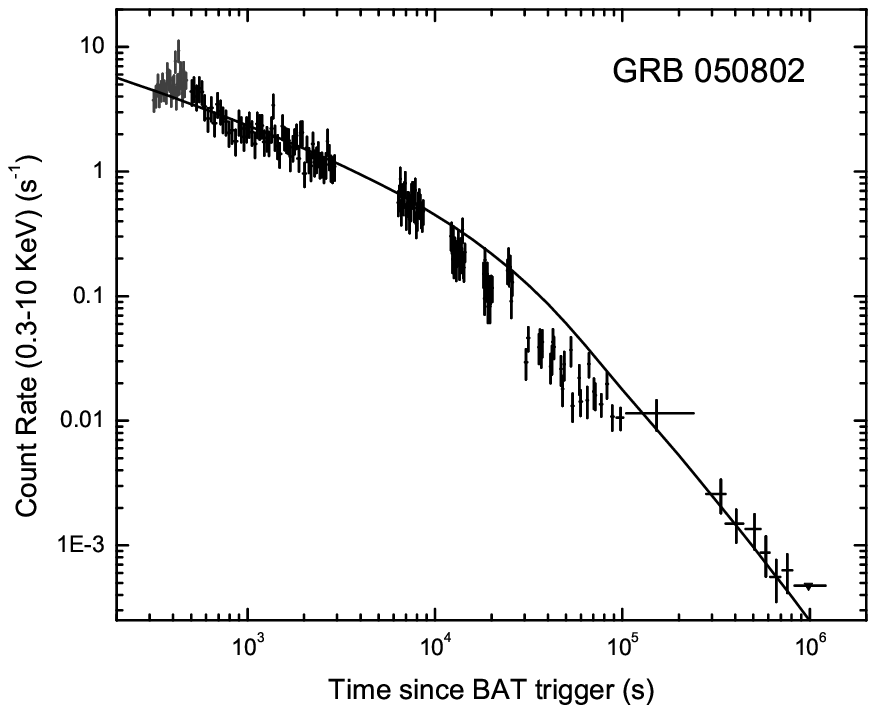}
\end{figure}
\begin{figure}
\plottwo{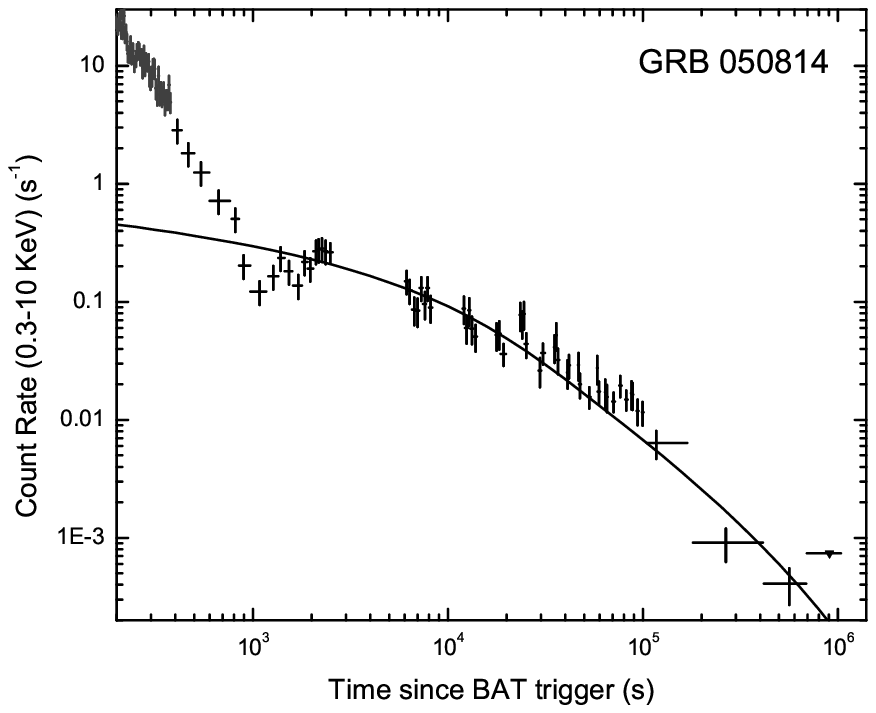}{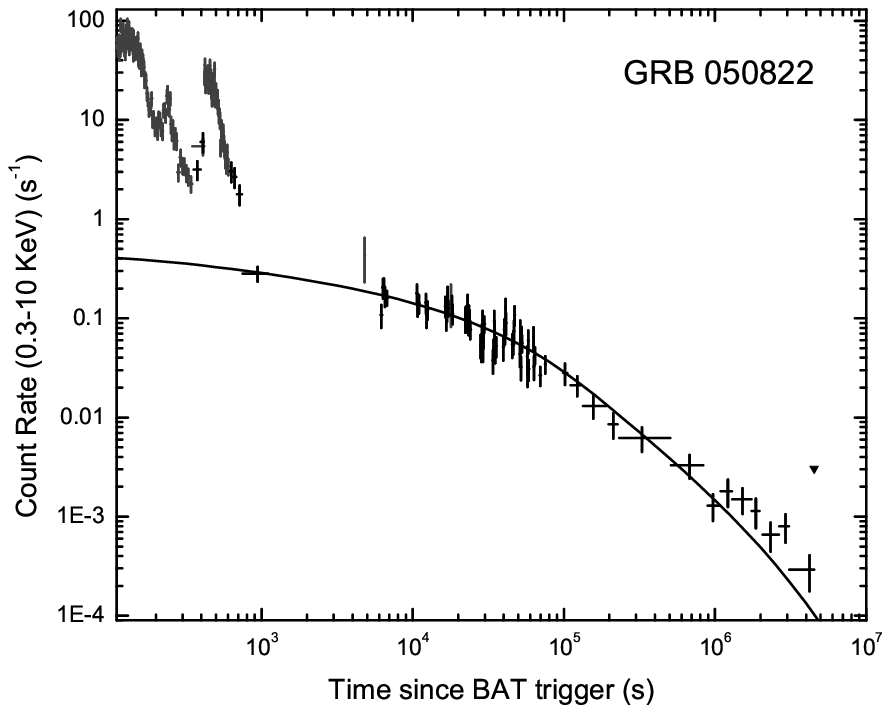}
\plottwo{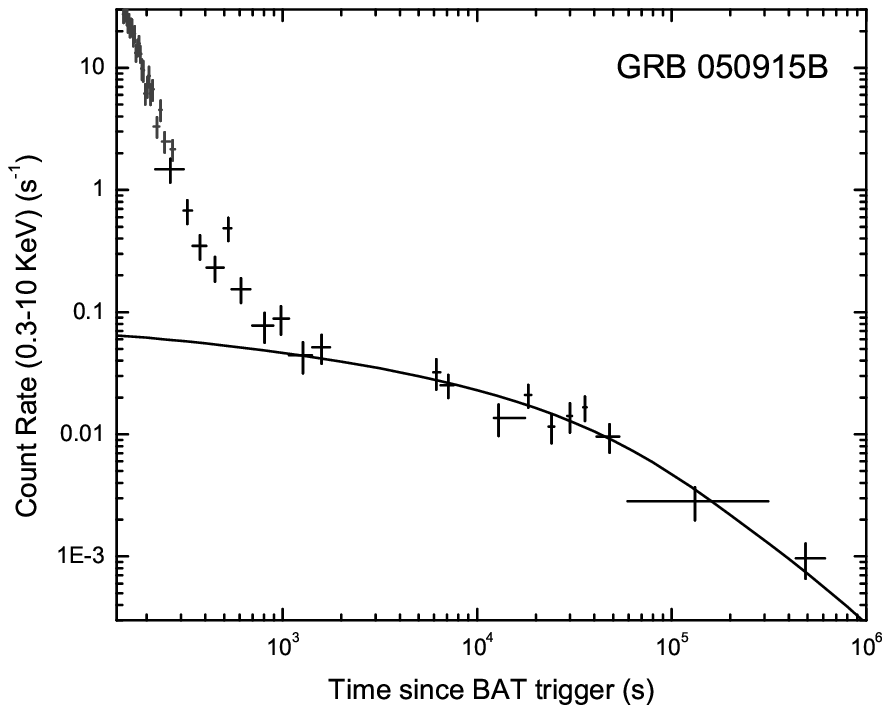}{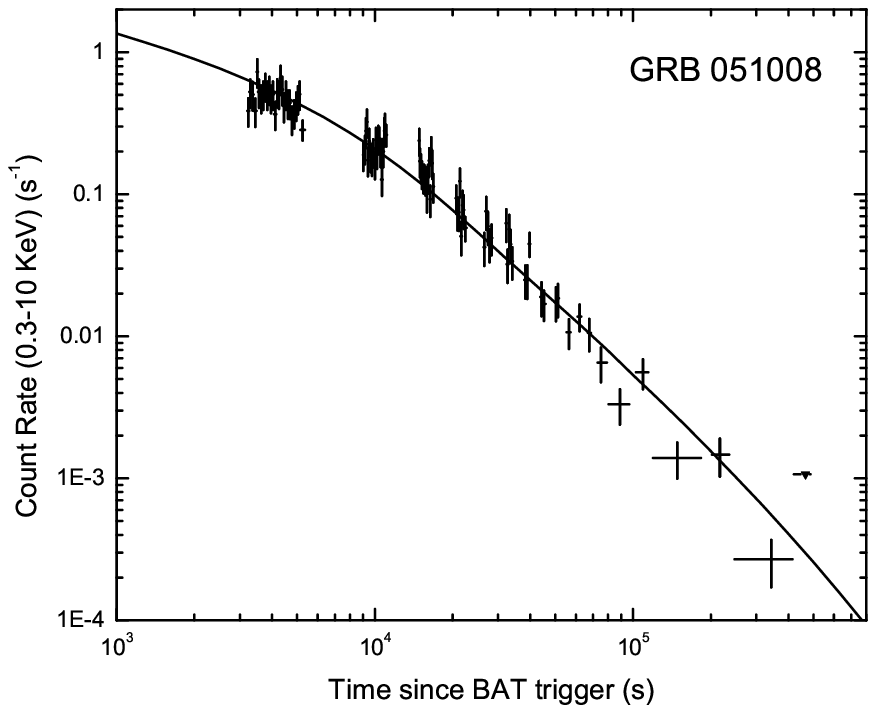}
\plottwo{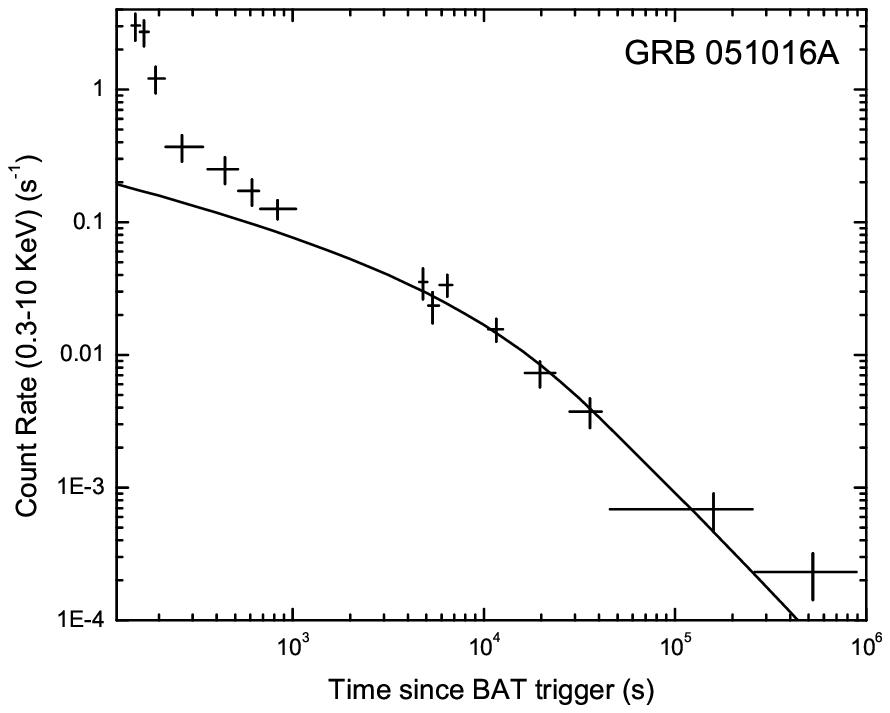}{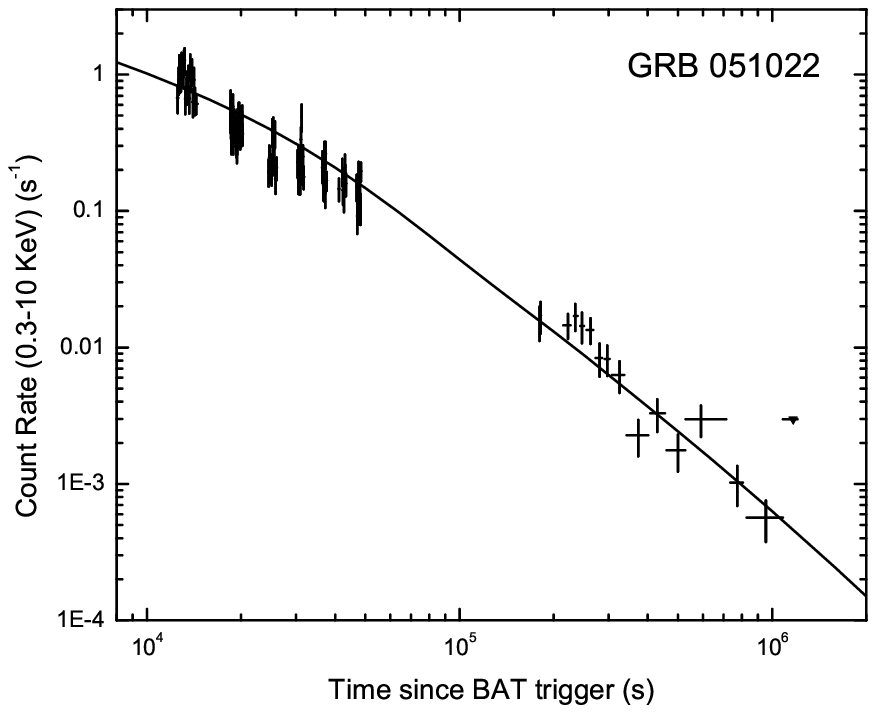}
\end{figure}
\begin{figure}
\plottwo{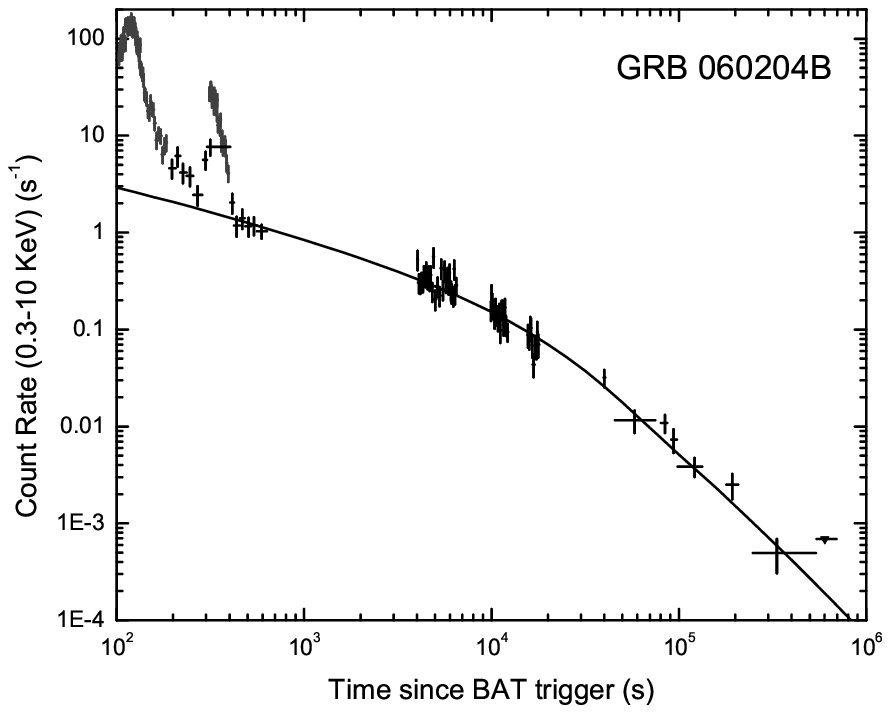}{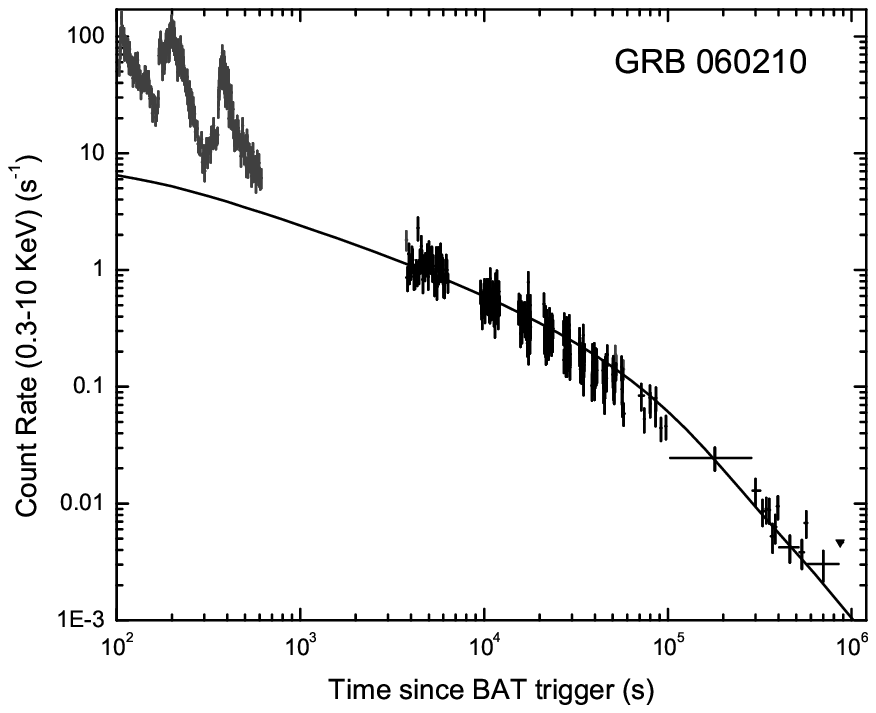}
\plottwo{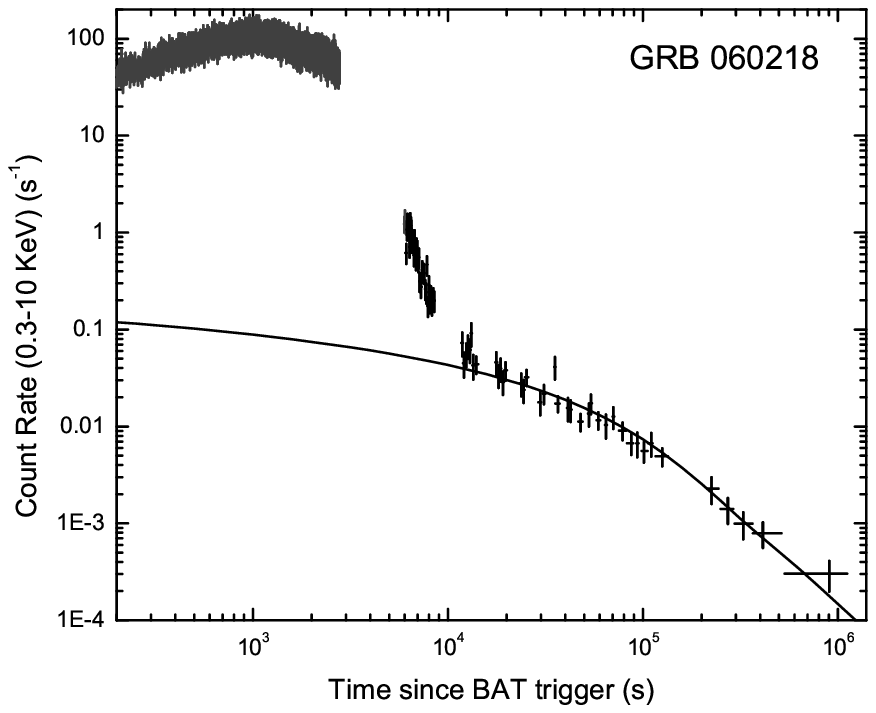}{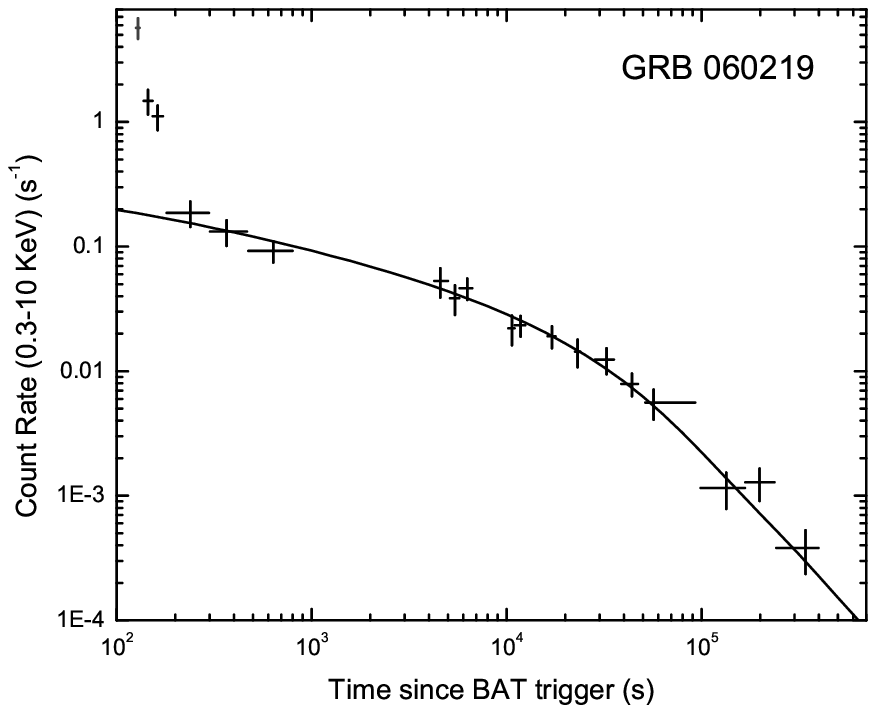}
\plottwo{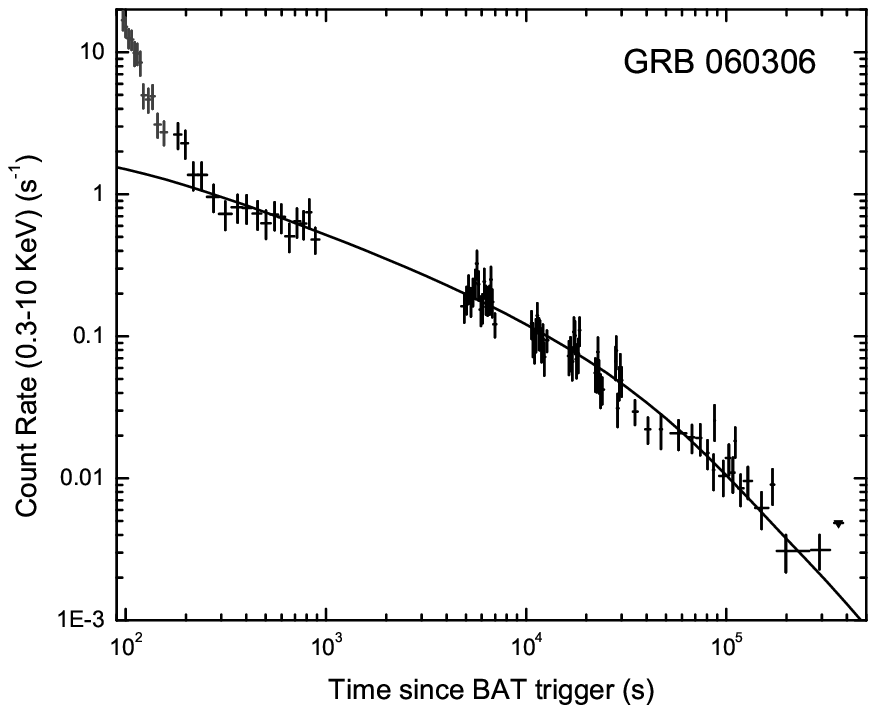}{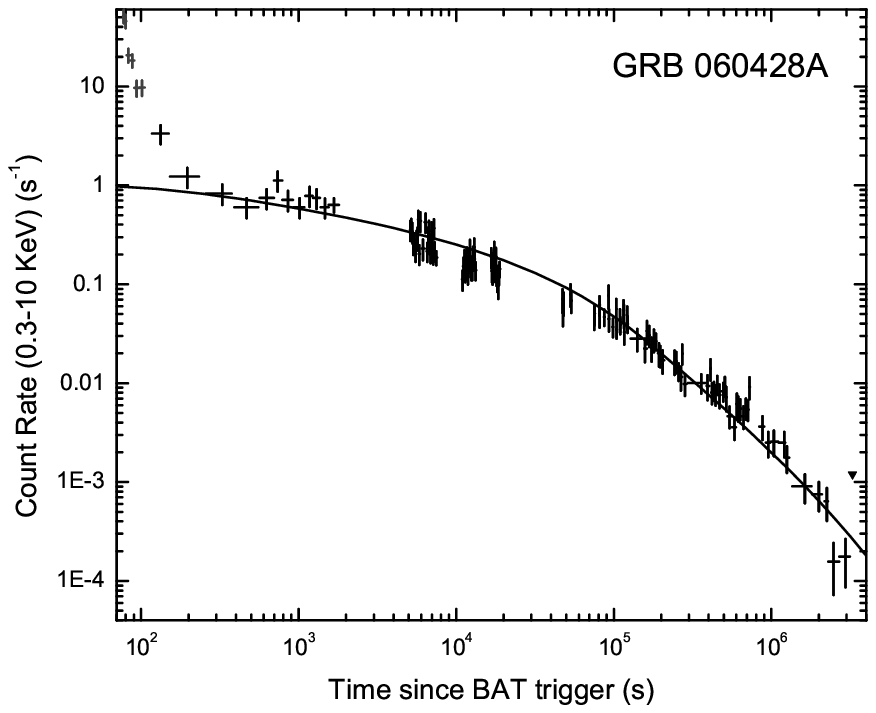}
\end{figure}
\begin{figure}
\plottwo{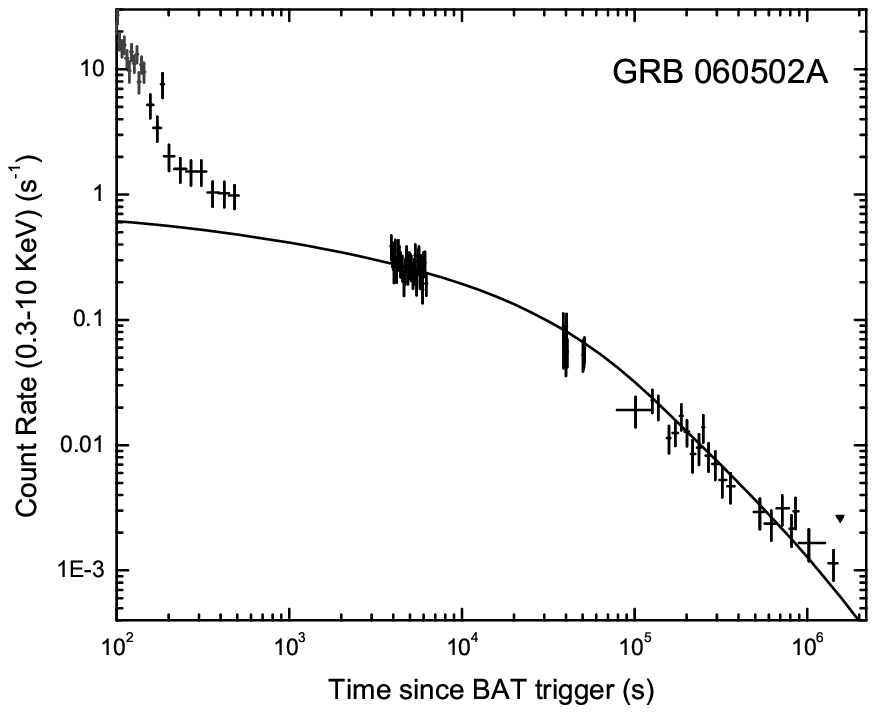}{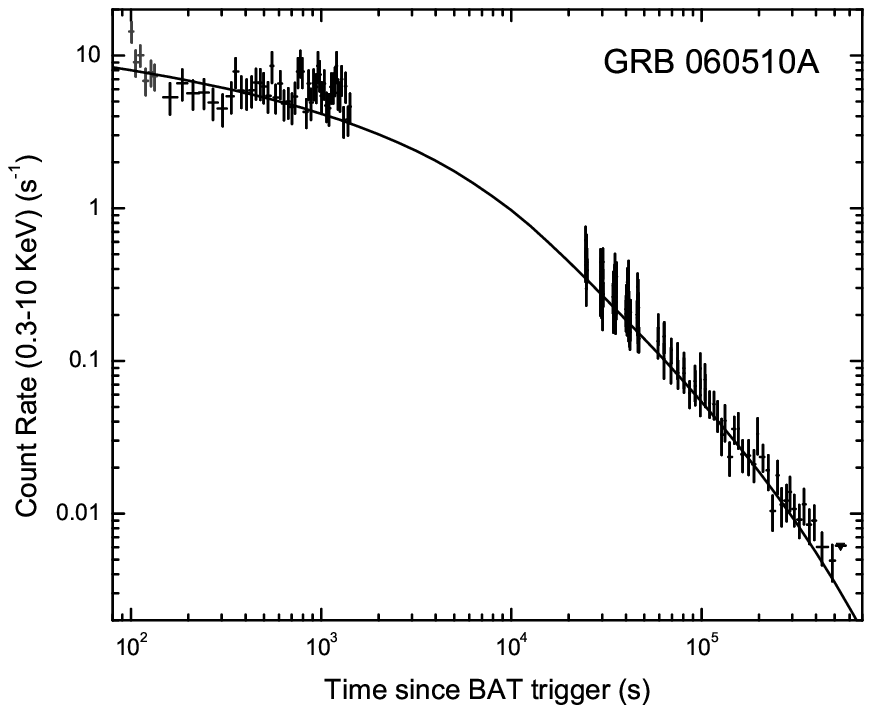}
\plottwo{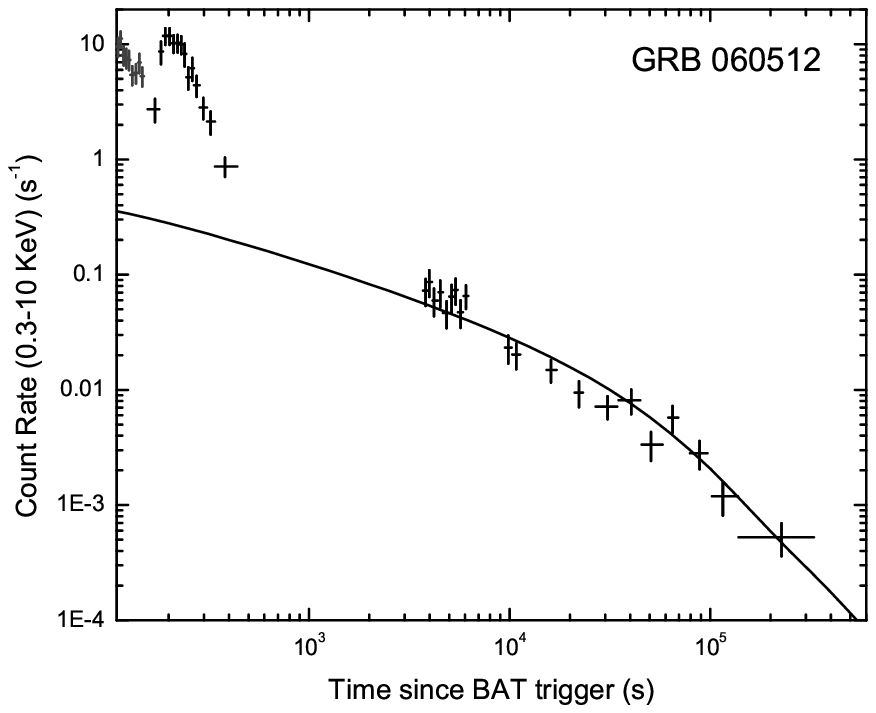}{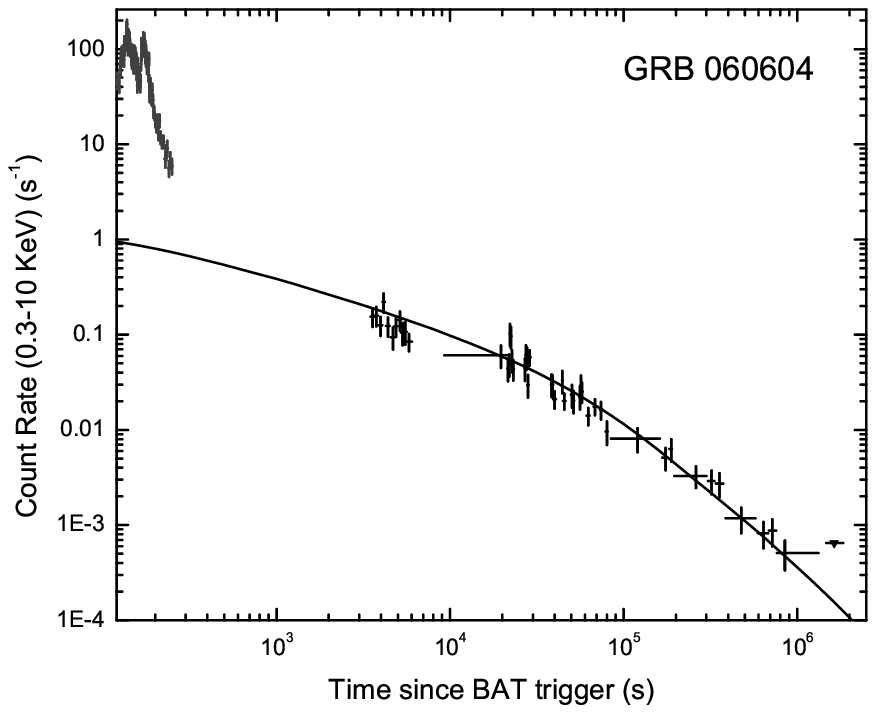}
\plottwo{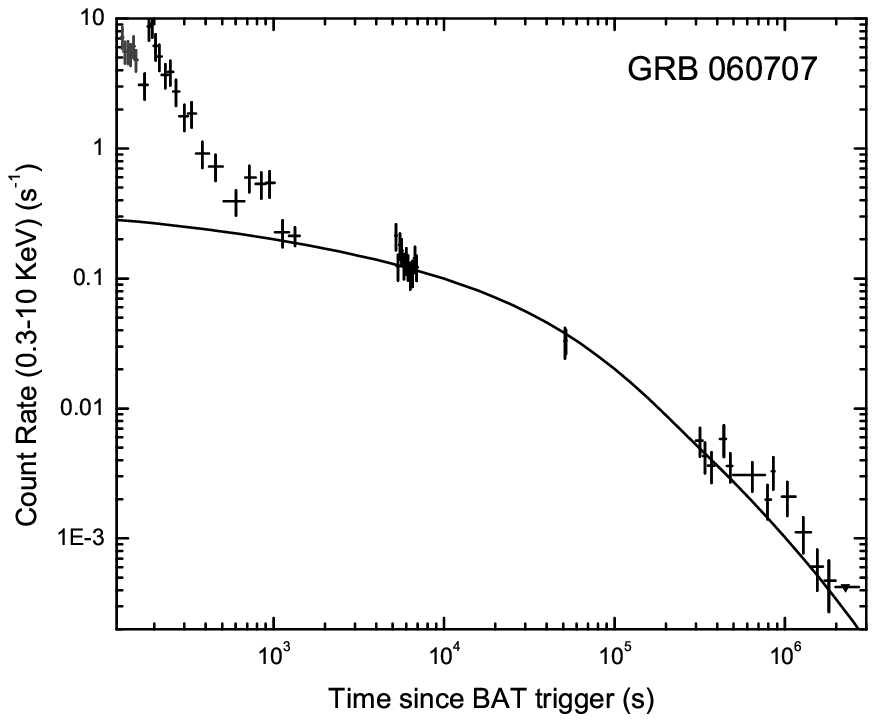}{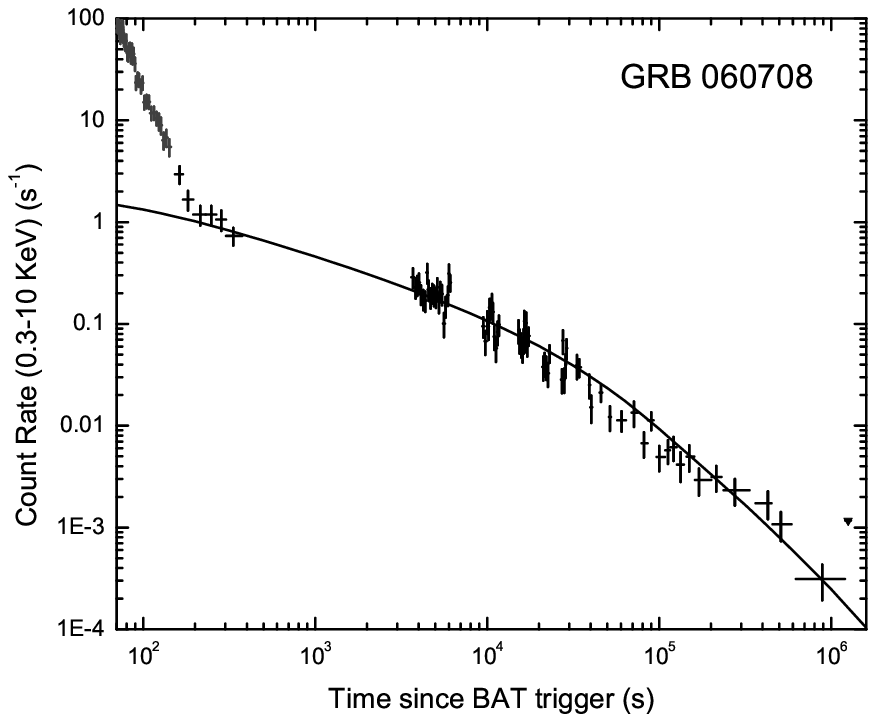}
\end{figure}
\begin{figure}
\plottwo{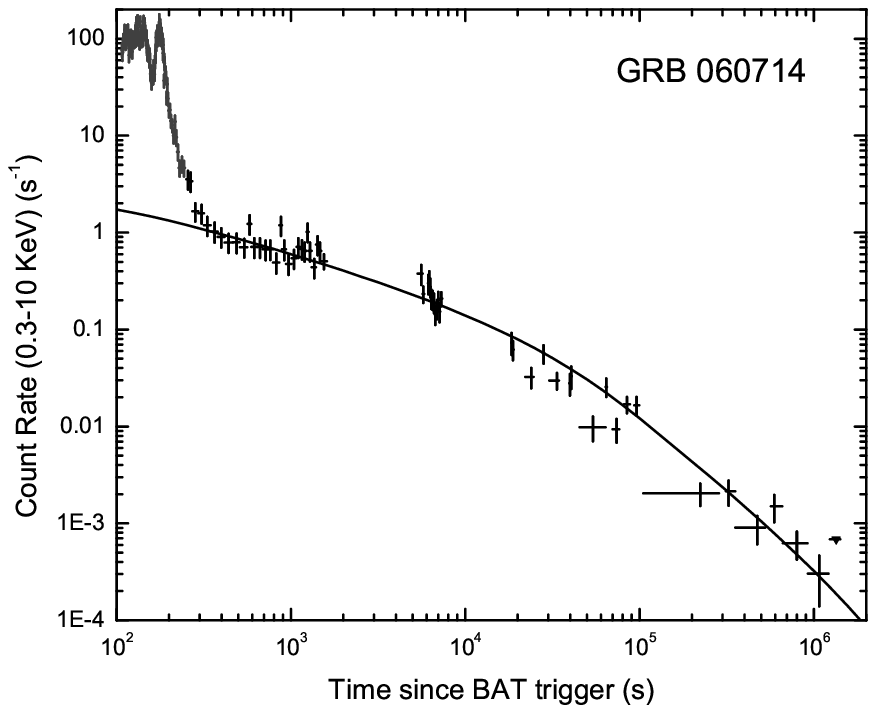}{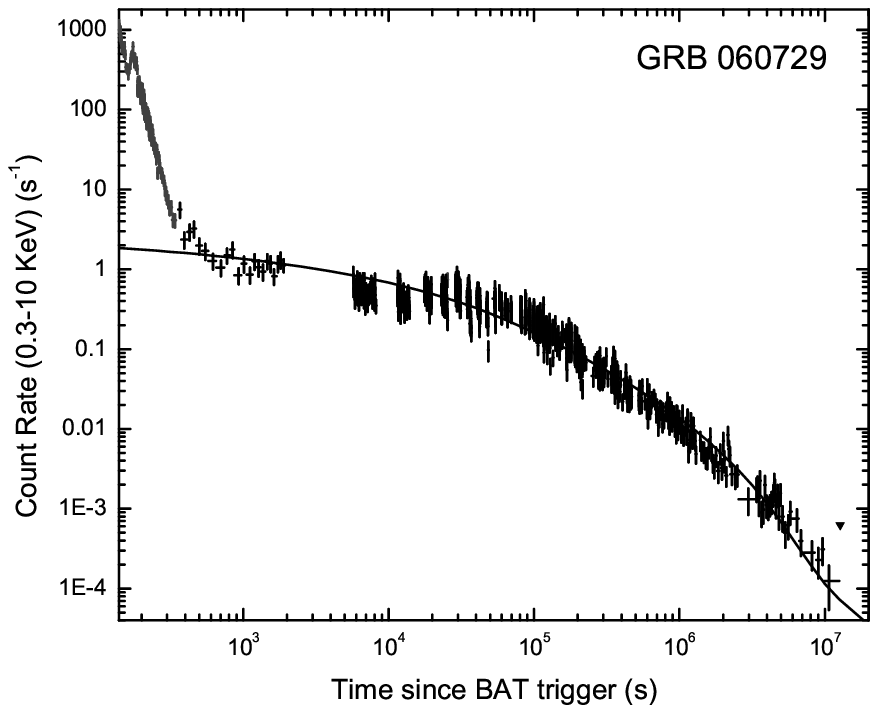}
\plottwo{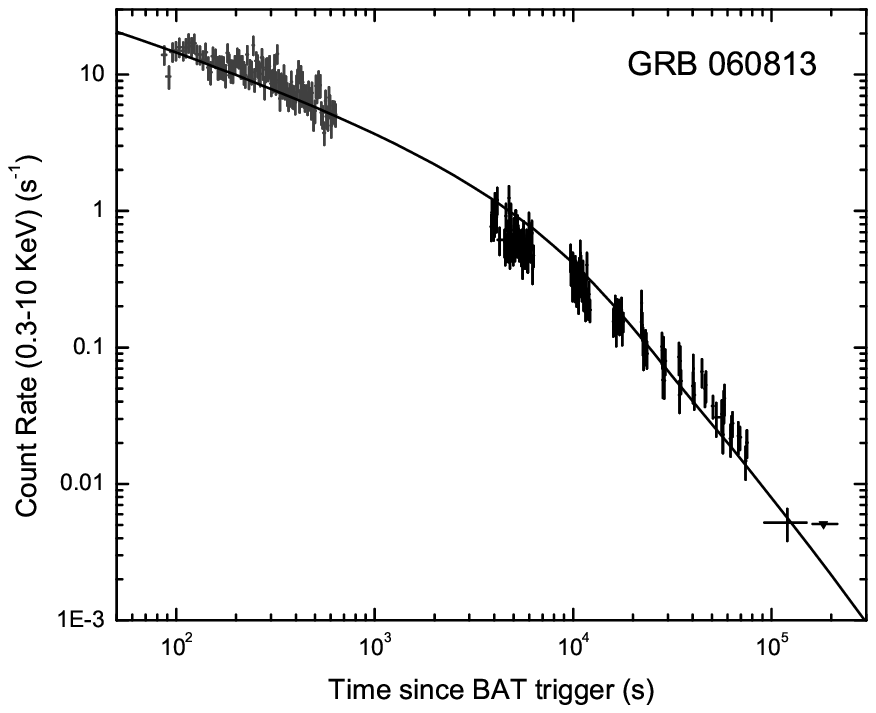}{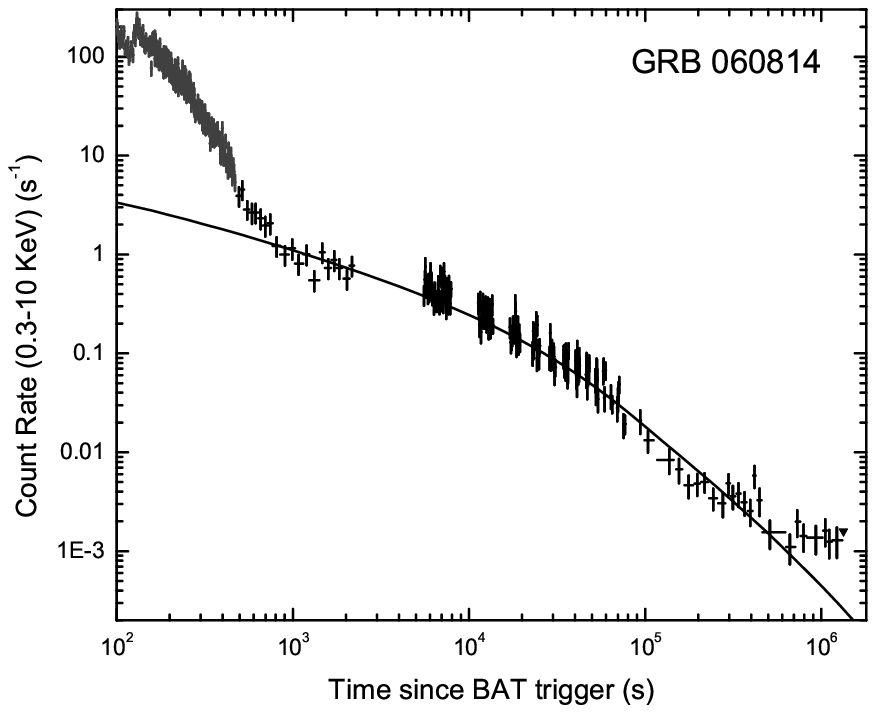}
\plottwo{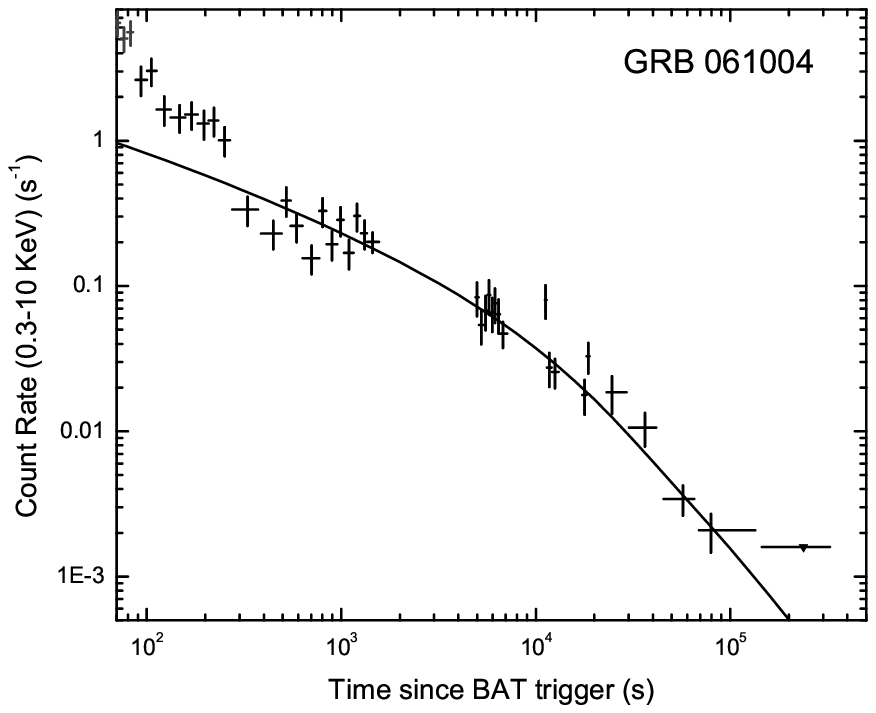}{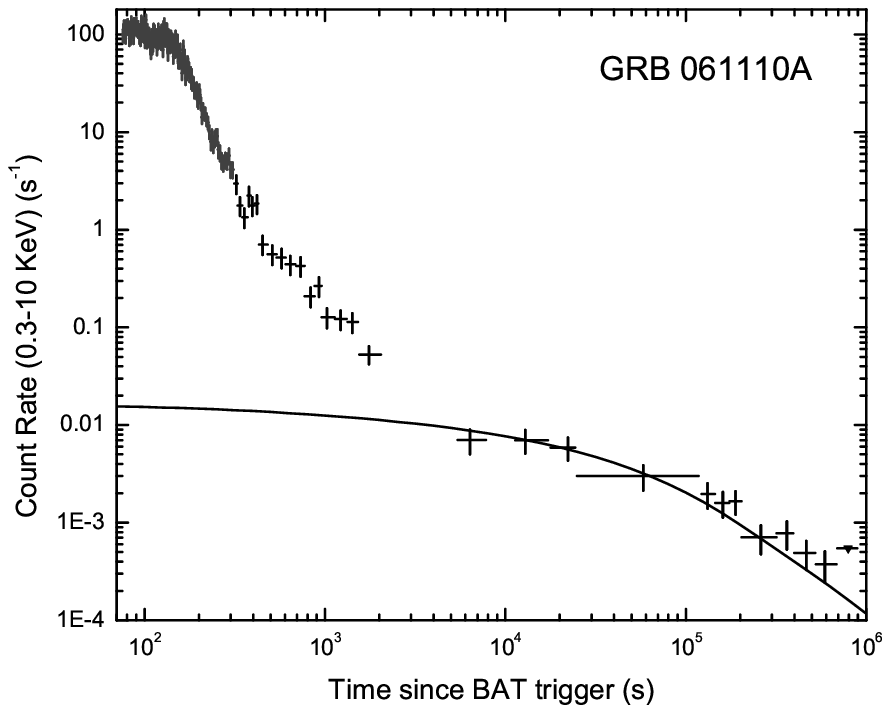}
\end{figure}
\begin{figure}
\plottwo{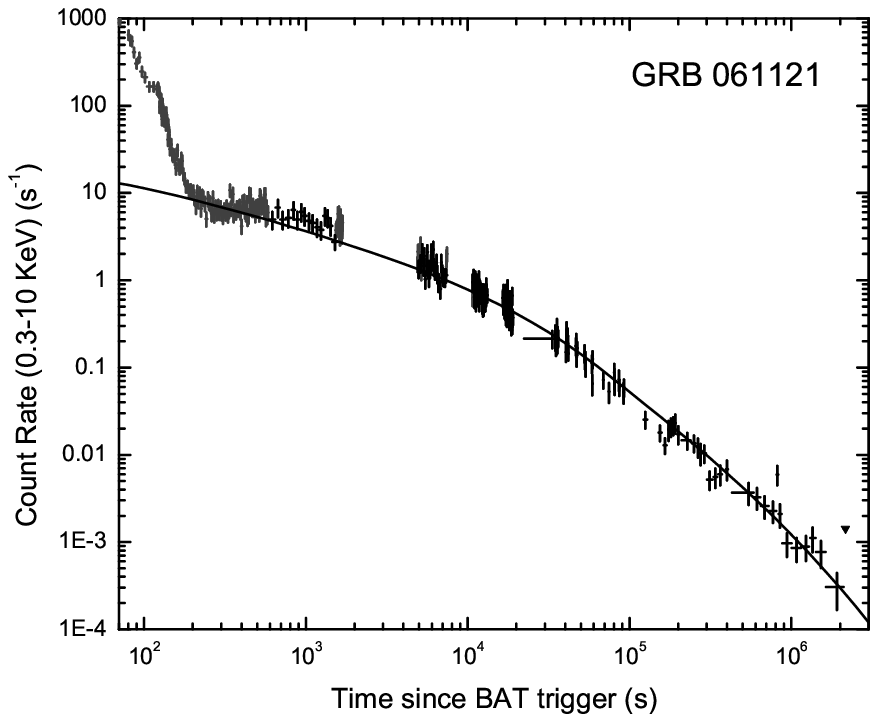}{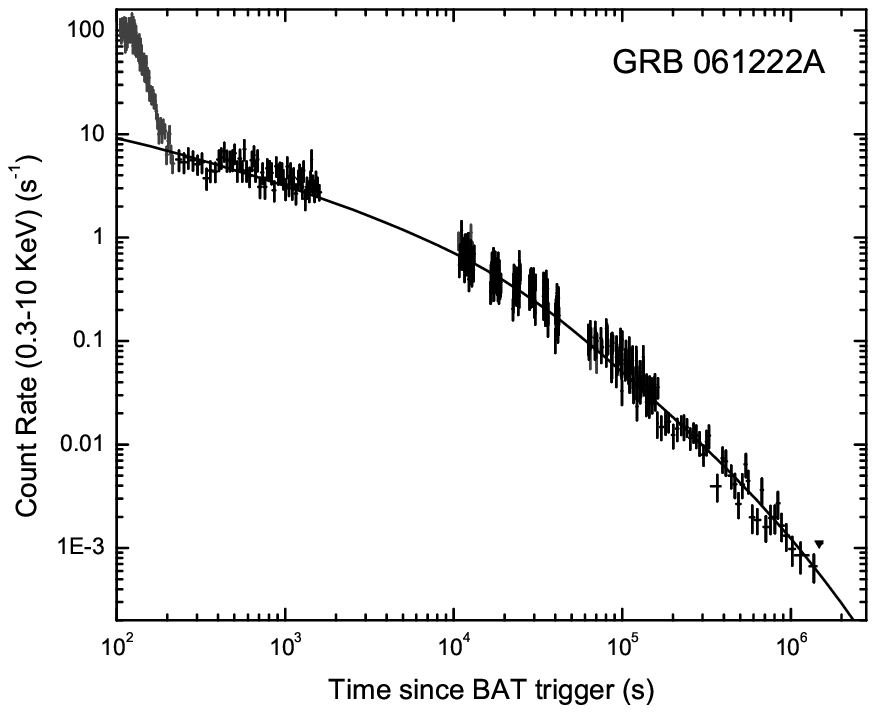} \plottwo{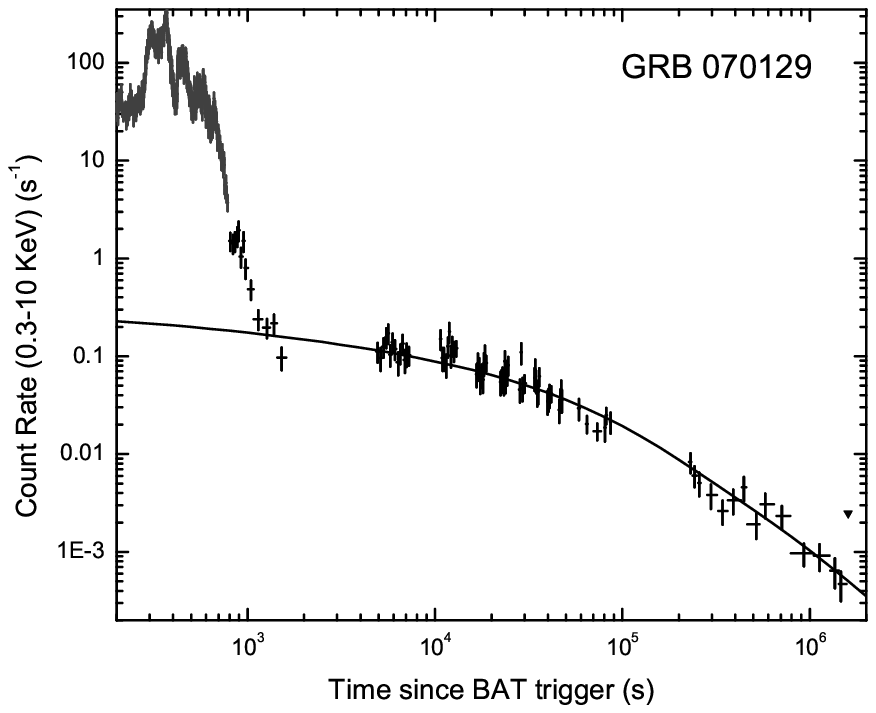}{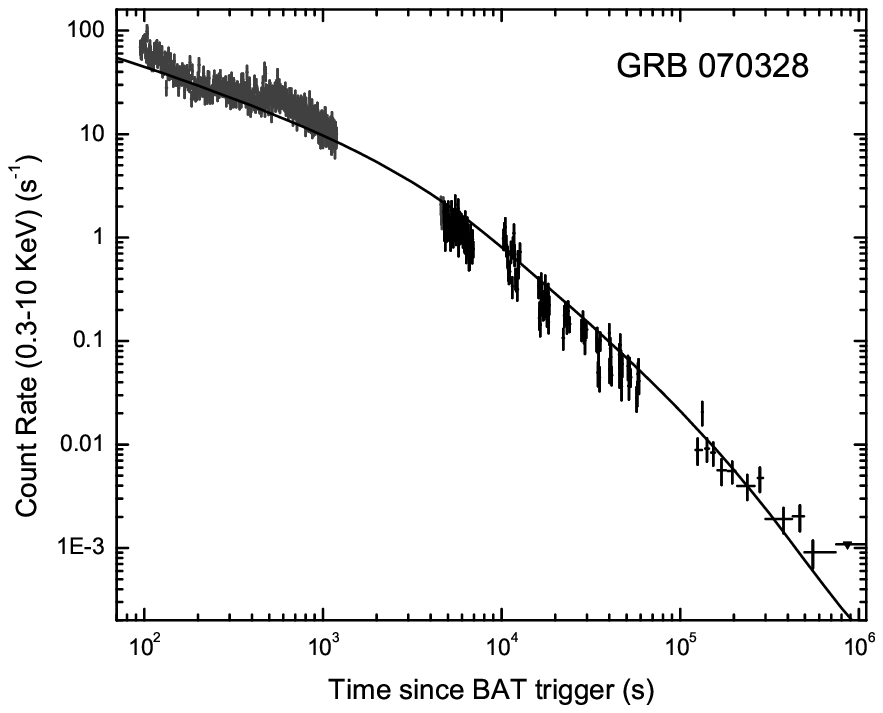}
\plottwo{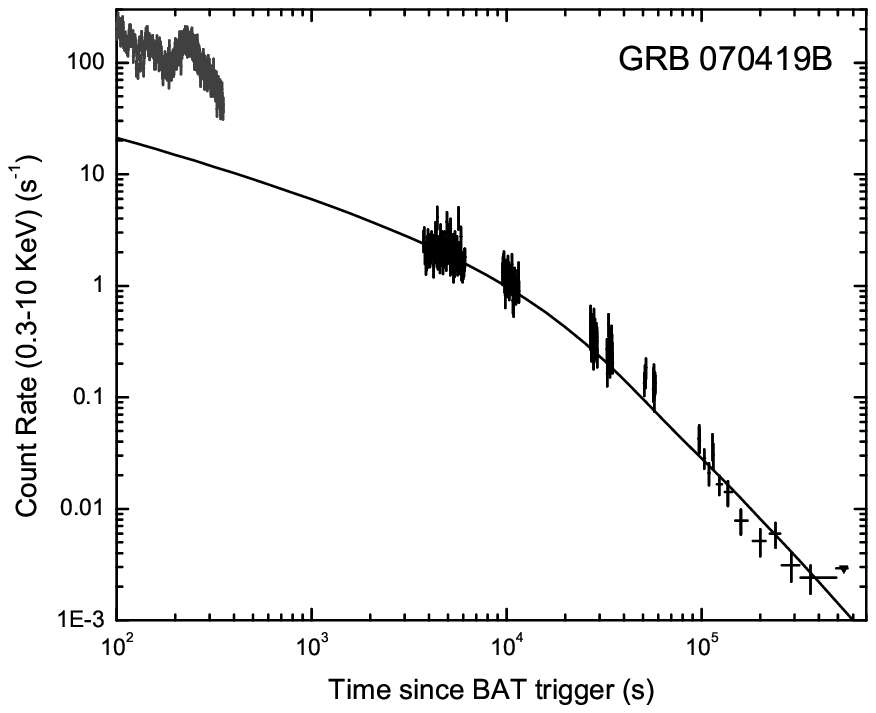}{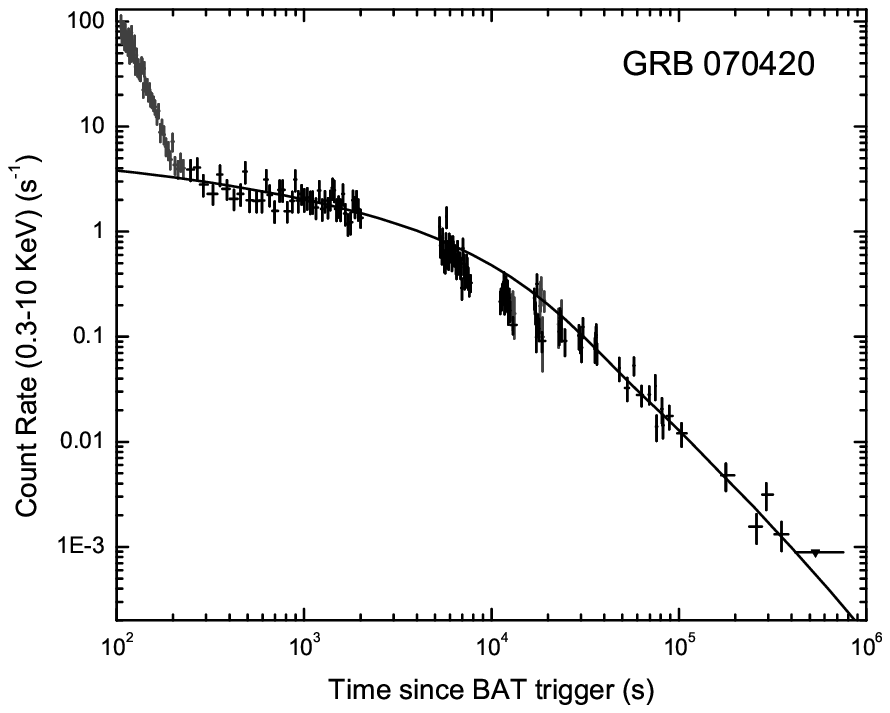} \caption{The X-ray light curves of 36
GRBs observed by \textit{Swift} XRT. WT mode is in gray, PC mode is
in black. The solid lines are the theoretical light curves given by
our model (Eq. \ref{eq:lightcurve}). We have $E_-=0.3$ keV and
$E_+=10$ keV for \textit{Swift} XRT, and we assume $a_-=0.005\,
\micron $, $\delta=0$, $E_{\rm p}=200$ keV. The other parameters,
$s$, $q$, $a_+$, and $R$ are given in Tab. \ref{tab:redshift} and
Tab. \ref{tab:noredshift}.} \label{fig:lightcurve}
\end{figure}

\begin{figure}
\plotone{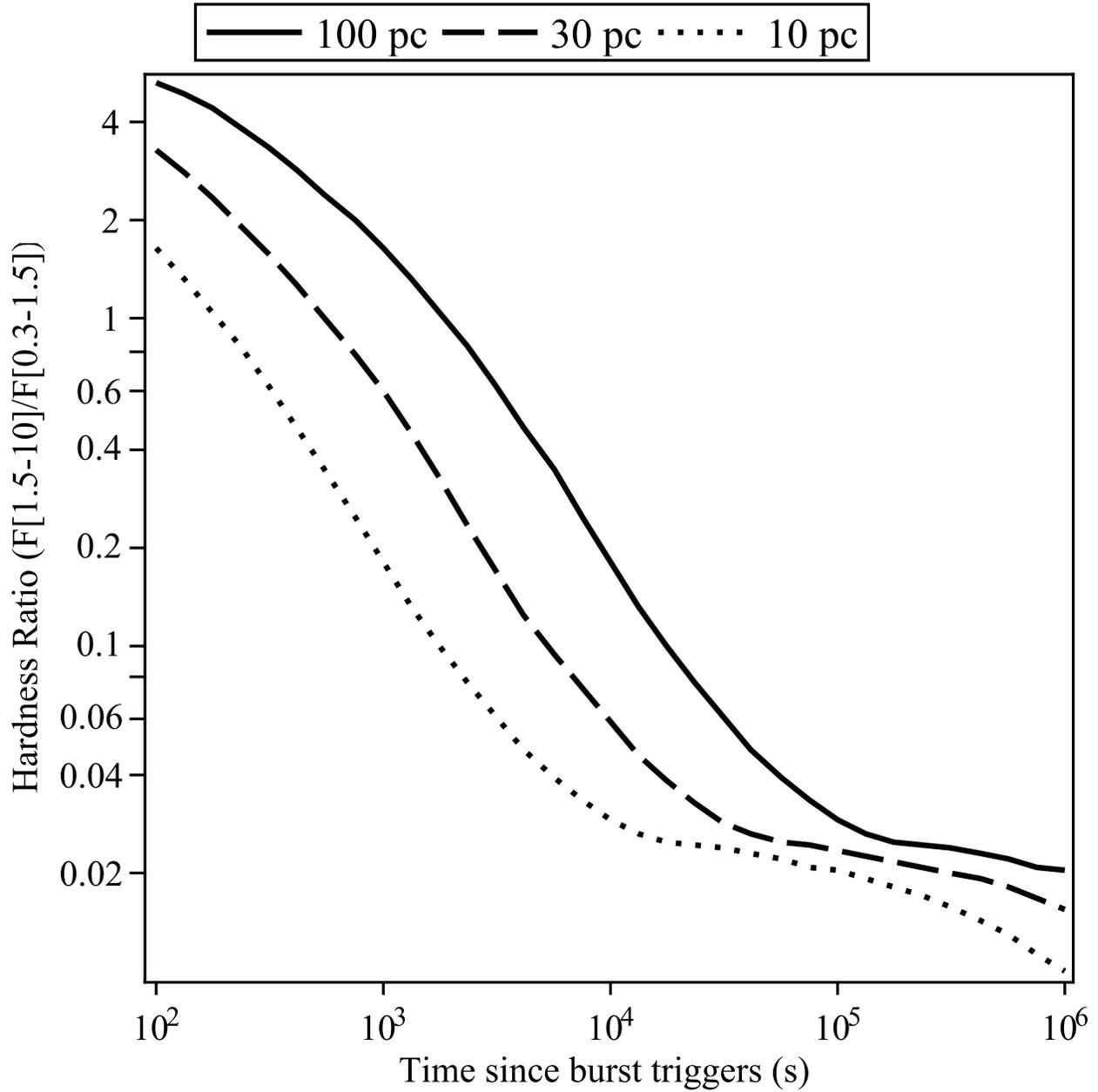} \caption{Evolution of the hardness ratio of echo
emission with different shell distance $R$. We assume $z=1$, $q=3.5$
and $a_+=0.025\, \micron$. Other parameters are the same with Fig.
\ref{fig:lightcurve}. The dramatically small ratio could be
alleviated only if the absorption in 0.3-1.5 keV is considered.}
\label{fig:hr}
\end{figure}

\end{document}